\documentclass[aps,reprint,twocolumn,superscriptaddress,nofootinbib,longbibliography,nobalancelastpage]{revtex4-1}
\usepackage{graphicx}
\usepackage{amssymb}
\usepackage{amsmath}
\usepackage{amssymb}
\usepackage{amsfonts}
\usepackage{bm}
\usepackage{braket}
\usepackage[dvipsnames]{xcolor}
\usepackage[colorlinks=True,citecolor=blue,urlcolor=blue,linkcolor=red]{hyperref}
\usepackage{hypcap}
\newcommand{\up}{\uparrow}
\newcommand{\dn}{\downarrow}

\newcommand{\gr}{{GOEastREM}}
\newcommand{\er}{{EastREM}}
\newcommand{\her}{H_\mathrm{EastREM}}
\newcommand{\hqr}{H_\mathrm{QREM}}
\newcommand{\hunconstrained}{\ensuremath{H_{\mathrm{X}}}}
\newcommand{\hconstrained}{\ensuremath{H_{\mathrm{East}}}}
\renewcommand{\o}{\mathcal{O}}
\newcommand{\wfs}{W_\mathrm{FS}}
\newcommand{\nh}{N_\mathcal{H}}
\newcommand{\np}{n^{(p)}}
\newcommand{\ns}{n^{(s)}}
\newcommand{\hd}{H_\mathrm{REM}}
\newcommand{\E}{\mathcal{E}}

\begin{document}
\begin{abstract}
  Quantum systems that violate the eigenstate thermalisation hypothesis thereby falling outside the paradigm of conventional statistical mechanics are of both intellectual and practical interest. We show that such a breaking of ergodicity may arise purely due to local constraints on random many-body Hamiltonians. As an example, we study an ergodic quantum spin-1/2 model which acquires a localised phase upon addition of East-type constraints. We establish its phenomenology using spectral and dynamical properties obtained by exact diagonalisation. Mapping the Hamiltonian to a disordered hopping problem on the Fock space graph we find that potentially non-resonant bottlenecks in the Fock-space dynamics, caused by spatially local segments of frozen spins, lie at the root of localisation. We support this picture by introducing and solving numerically a class of random matrix models that retain the bottlenecks. Finally, we obtain analytical insight into the origins of localisation using the forward-scattering approximation. A numerical treatment of the forward-scattering approximation yields critical points which agree quantitatively with the exact diagonalisation results.
\end{abstract}

\title{Strong ergodicity breaking due to local constraints in a quantum system}

\author{Sthitadhi Roy}
\email{sthitadhi.roy@chem.ox.ac.uk} 
\affiliation{Rudolf Peierls Centre
  for Theoretical Physics, Clarendon Laboratory, Oxford University,
  Parks Road, Oxford OX1 3PU, United Kingdom} 
\affiliation{Physical
  and Theoretical Chemistry, Oxford University, South Parks Road,
  Oxford OX1 3QZ, United Kingdom} 

\author{Achilleas Lazarides}
\email{A.Lazarides@lboro.ac.uk} 
\affiliation{Interdisciplinary Centre
  for Mathematical Modelling and Department of Mathematical Sciences,
  Loughborough University, Loughborough, Leicestershire LE11 3TU,
  United Kingdom}
\maketitle

\section{Introduction}

\label{sec:introduction}

The fundamental question of how closed quantum systems admit a
thermodynamic description has motivated the study of their excited
eigenstates and out-of-equilibrium
dynamics~\cite{srednicki1999approach,rigol2008thermalization}.  A
central concept is the Eigenstate Thermalisation
Hypothesis (ETH), satisfied by generic ergodic
systems~\cite{deutsch1991quantum,srednicki1994chaos,dalessio2016quantum,deutsch2018eigenstate}.
Positing that eigenstate expectation values of local observables are
smooth functions of the eigenenergies, ETH amounts to a statement
that the energy, an integral of motion, is a state variable so that
local observables in the long-time dynamical state are fully
determined by its value.\footnote{Assuming no other conserved
  quantities, for example, total spin, momentum, particle
  number etc.} Generic systems satisfy this by default, and any
violation of the ETH is therefore interesting.  In a recent
development it was realised that one way ETH can be violated is the
presence of many-body localisation
(MBL)~\cite{basko2006metal,gornyi2005interacting,oganesyan2007localisation,znidaric2008many,pal2010many}
(see
Refs.~\cite{nandkishore2015many,alet2018many,abanin2019colloquium} for
reviews and further references therein).  Specifying macroscopic
properties of MBL systems requires an extensive set of emergent
quasi-local integrals of
motion~\cite{serbyn2013local,huse2014phenomenology,ros2015integrals,rademaker2016explicit,imbrie2017local}.

\begin{figure}[t]
  \includegraphics[width=\columnwidth]{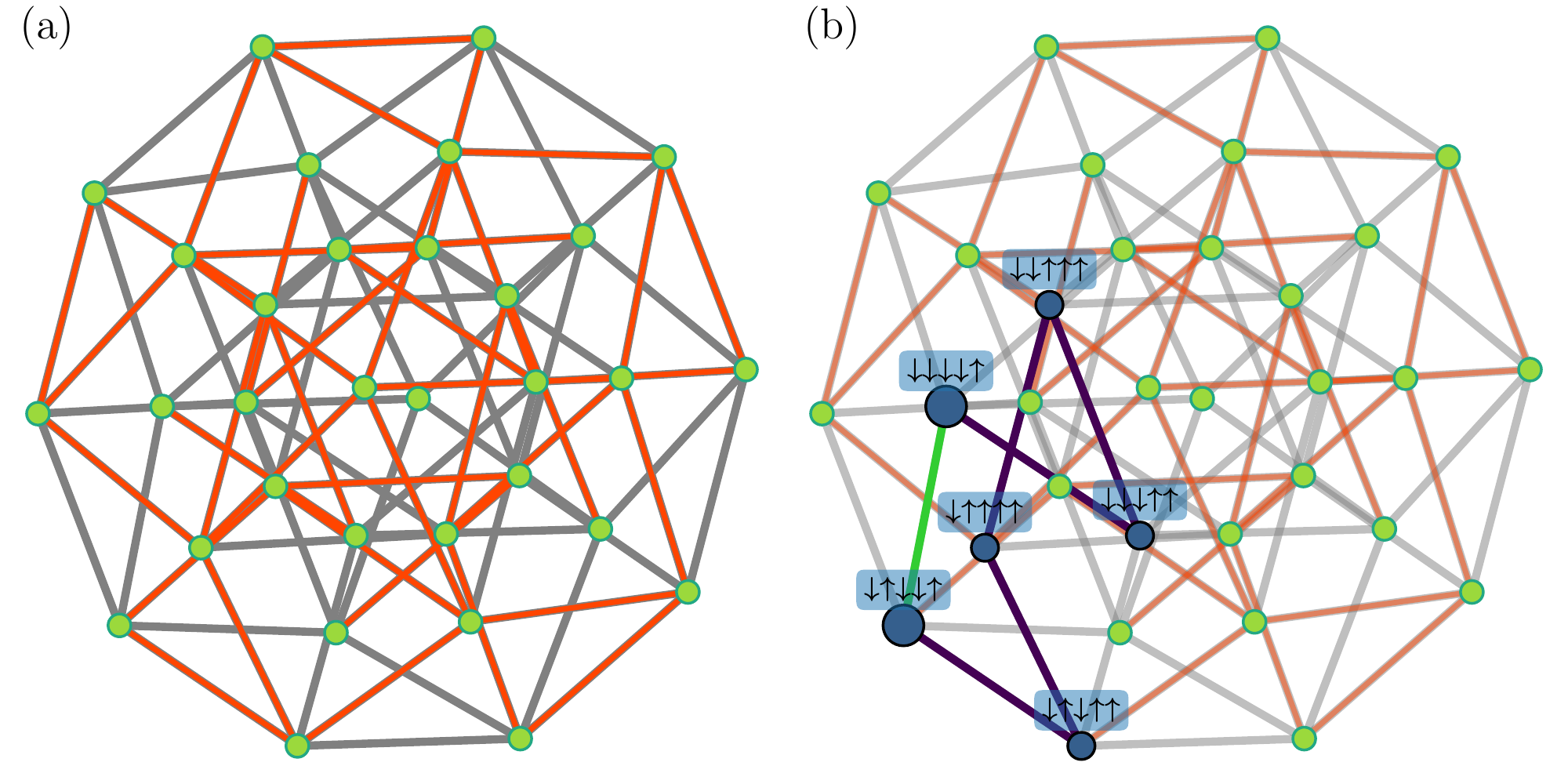}
  \caption{The Fock-space graph for a system of $N (= 5)$ spins-1/2
     where each vertex is a $\sigma^z$-product state as indicated and
    the links correspond to spin-flip terms in the Hamiltonian. (a)
    The links in red are present for both the QREM as well as for the
    EastREM whereas the those in grey are present only for the QREM;
    the constraints in the EastREM switches off the latter. (b)
    Illustration of how the constraints generally increase the
    shortest distance between two vertices. For the pair of vertices
    chosen, the path with the QREM is only of length one (green)
    whereas for the EastREM the shortest path shown in black is
    longer.}
  \label{fig:eastrem-vs-qrem}
\end{figure}

Many of the universal properties of ETH systems are well described
within the framework of random matrix theory~\cite{deutsch1991quantum,srednicki1994chaos,dalessio2016quantum,deutsch2018eigenstate,mehta2004random}, where one studies random
matrices incorporating the relevant symmetries of the physical system
instead of actual Hamiltonians.
The physical intuition here is that a random matrix is the ``least
structured'' object which still captures the characteristic properties
of the Hamiltonians of interest. Inspired by this approach, and
focussing on Hamiltonians with finite local Hilbert space dimensions
(spin-1/2 systems constituting possibly the simplest example), we ask
what properties a random many-body Hamiltonian matrix must satisfy so
as to capture the defining properties of MBL systems, namely,
violation of ETH and absence of spatial transport.

As the question pertains to the many-body Hamiltonian, it is natural
to treat the problem directly as one of localisation in Fock
space~~\cite{logan1990quantum,altshuler1997quasiparticle,logan2019many,roy2019self,pietracaprina2016forward,pietracaprina2019hilbert,roy2018exact,roy2018percolation,detomasi2019dynamics,roy2019fock}.
Any many-body Hamiltonian of our type can be interpreted as a
disordered hopping problem on the Fock space of the
system. Considering the Fock basis states as vertices and the hoppings
as links between them, one can view the Hamiltonian matrix as a graph
(see Fig.~\ref{fig:eastrem-vs-qrem}).\footnote{The adjacency matrix of
  this graph is closely related to the Hamiltonian expressed in the
  Fock basis.} The question then translates to what properties the
graph must have for the system to exhibit non-ergodic behaviour.

In this paper we provide one answer to the question: \emph{constrained
  hopping on the Fock space due to local kinetic constraints on the
  real-space degrees of freedom can lead to localisation}. 
Such constrained dynamics in
translation invariant systems have been to shown to exhibit slow
dynamics and metastable
behaviour~\cite{vanhorssen2015dynamics,lan2018quantum,pancotti2019quantum}, but here we focus on the possibility of a localised phase and accompanying localisation transitions in the eigenstates.
Using exact numerical and approximate analytical techniques we demonstrate that
constraining Fock-space connectivity leads to localisation, and that
this is not due to disorder as the unconstrained version of our model
is disordered but never localised. The Fock space of our model is not
fragmented but is rather fully connected, thus the physics here is
qualitatively different from cases where the constraints fragment Fock
space~\cite{pai2019localization,khemani2019local,sala2019ergodicity}. Furthermore
our model breaks ergodicity strongly, as signified by the presence of a
phase where all eigenstates are localised, thus also differing from
weak ergodicity breaking as in the case of quantum many-body
scars~\cite{turner2018weak,turner2018quantum,khemani2019signatures}. In our case, ergodicity
breaking is due to the states in Fock space naturally grouping into
clusters, with dense intra-cluster but sparse inter-cluster
connections. This leads to potentially non-resonant bottlenecks in the
Fock space, which is the root of localisation.
This establishes the central result of this work -- \emph{how
  constrained quantum dynamics can lead to localisation without
  fragmenting the Fock space}.

\subsection*{Structure of the paper}

We start with an overview of the paper in Sec.~\ref{sec:overview}
which sets up the Fock space we work with and states the main results
of this paper.  In Sec.~\ref{sec:model} we introduce a kinetically
constrained quantum spin-1/2 model to put the ideas on a concrete
footing.  The constrained quantum model is based on the quantum random
energy model (QREM) which acts as our reference unconstrained
model. The QREM has been shown to be completely delocalised except for
a vanishing fraction of eigenstates at the spectral
edges~\cite{goldschmidt1990solvable,laumann2014many,baldwin2016manybody,baldwin2017clustering}. We
then impose East model-like
constraints~\cite{ritort2003glassy,garrahan2011kinetically,vanhorssen2015dynamics,garrahan2018aspects},
calling the resulting model the \er. Section~\ref{sec:phenomenology}
is dedicated to the phenomenology of the model: we map out its phase
diagram using spectral and eigenstate properties in
Sec.~\ref{subsec:spectral}, finding that a fully localised phase
emerges, unlike for the QREM. Dynamical properties further support
this as shown in Sec.~\ref{subsec:dynamics} where we find that an
initial state retains its memory locally in space, reflecting the
locality of the constraints. In Sec.~\ref{subsec:clustering} we
discuss how the constraints impose a particular structure in the
Hamiltonian and construct Hamiltonians random apart from having this
structure, showing that they still display the dynamics of
interest. Finally, in Sec.~\ref{sec:fsa} we use the forward scattering
approximation (FSA) to obtain a (semi)-analytical understanding of
localisation in this model: Secs.~\ref{subsec:bulk} and
\ref{subsec:edges} present an analytical treatment of the FSA for the
spectral bulk and edges respectively, while
Sec.~\ref{subsec:numerical-fsa} presents a numerical treatment of the
FSA, finding agreement with the numerical results of
Sec.~\ref{sec:phenomenology}. The FSA explicitly demonstrates the role
of constraints and reveals clearly the distinction between the
unconstrained (QREM) and constrained (\er) versions of the model,
explicitly demonstrating the role of the constraints.

\section{Overview}
\label{sec:overview}
Fock space offers a natural viewpoint from which to approach the
problem as any many-body Hamiltonian can be interpreted as a hopping
problem on the Fock-space graph,
\begin{equation}
  H=\underbrace{\sum_{\alpha=1}^{\nh} \E_\alpha\ket{\alpha}\bra{\alpha}}_{H_\mathrm{diag}} + \underbrace{{\sum_{\alpha\neq\beta}}\Gamma_{\alpha\beta} \ket{\alpha}\bra{\beta}}_{H_\mathrm{offdiag}}.
  \label{eq:ham-FS}
\end{equation}
Here the set of basis states $\{\ket{\alpha}\}$ are the sites on the
$\nh$-dimensional Fock-space graph, of which there are exponentially
many (in system size), $\nh\sim e^N$.  The diagonal elements of the
Hamiltonian $\E_\alpha$ are the on-site energies in this
Fock space.  The off-diagonal elements $\Gamma_{\alpha\beta}$ then
represent hopping amplitudes. The offdiagonal part of the Hamiltonian,
$H_\mathrm{offdiag}$ also allows us to define a distance on the Fock
space graph -- the distance between two states $\ket{\alpha}$ and
$\ket{\beta},$ denoted as $r_{\alpha\beta}$, is defined as the length of
the shortest path between them following the links generated by
$H_\mathrm{offdiag}$.

Hamiltonian matrices as in Eq.~\eqref{eq:ham-FS} which are associated
with a many-body system (short-ranged with local degrees of freedom)
in general have matrix elements which satisfy two generic
features. First, the Fock-space site energies scale as $\sqrt{N}$,
such that one can define an effective on-site disorder strength on the
Fock space as $\wfs^2:=\nh^{-1}\sum_\alpha \braket{\E_\alpha^2}=W^2 N$
and $W$ an $\o(1)$ number. This simply reflects that for generic
short-ranged systems, each Fock-space site energy is an extensive sum
of random numbers.  Secondly, the off-diagonal matrix elements are
numbers of magnitude $\o(1)$ and, crucially, the average connectivity
of the Fock-space sites is extensive:
$\nh^{-1}\sum_{\alpha,\beta}\braket{\Gamma_{\alpha\beta}^2}\sim
N$. This is a result of the fact that for short-ranged systems, the
Hamiltonian connects a state with an extensive number of different
states each differing from the initial one only locally.

If all the Fock-space site energies are independent of each other,
Eq.~\eqref{eq:ham-FS} can be interpreted as an Anderson localisation
problem on a graph with connectivity $N$, hopping amplitude $\Gamma$,
and disorder strength $\wfs$.  Applying the localisation criterion for
Bethe lattices~\cite{abou-chacra1973self}, which we expect to work
well for cases with diverging connectivity, one finds the critical
disorder strength $W_c$ satisfies
$\frac{2e\sqrt{N}\Gamma}{W_c}\ln\left(\frac{W_c\sqrt{N}}{2\Gamma}\right)=1,$
so that $W_c$ diverges in the thermodynamic limit.\footnote{The expression is obtained from the so-called `upper limit approximation', however the exact solution was shown to differ from it by a factor $\approx e/2$~\cite{anderson1958absence,abou-chacra1973self}.} A localised phase
therefore does not exist, at least in the bulk of the spectrum.  We
therefore ask what additional ingredients are minimally required to
stabilise a many-body localised phase without altering the generic
features mentioned above.

Elsewhere~\cite{roy2019fock}, one answer to this question was
provided: strong correlations in the $\E_\alpha$, which render the
problem fundamentally different from an Anderson localisation problem
on a high-dimensional graph.\footnote{Here, strong correlations means
  that two basis states $\ket{\alpha}$ and $\ket{\beta}$ finitely
  distant from each other have Fock-space site energies finitely
  different from each other in the thermodynamic limit,
  $\vert \E_\alpha-\E_\beta\vert \sim\mathcal{O}(1)$.} In fact, this
is precisely the scenario for local Hamiltonians where the presence of
a localised phase has been argued for on analytical as well as
numerical
grounds~\cite{oganesyan2007localisation,pal2010many,luitz2015many,lev2015absence,imbrie2016many,logan2019many,roy2019fock}.

In this work, we take the complementary perspective and show that,
depending on the pattern and distribution of connectivities, a fully
localised phase may occur even for completely \emph{uncorrelated}
Fock-space disorder, \emph{non-fractured (i.e., fully-connected)} Fock
space, and \emph{typically extensive} connectivity for each site. We
demonstrate this for the case of spatially local kinetic constraints,
which create bottlenecks in the Fock space but leave it fully
connected (every site is accessible from every other).

Although our Fock space is not fragmented, it can be reorganised into
sparsely connected clusters. The picture that emerges is one of sites
densely interconnected within each cluster, but sparse
interconnections between clusters. In other words, the constraints
suppress links between sites belonging to different clusters. We show
that this is the fundamental mechanism which leads to a fully
many-body localised phase in both real and Fock spaces, despite the
Fock-space site energies being uncorrelated and the Fock space not
being fragmented -- this constitutes the central result of this work.

As a concrete setting we consider a system of $N$ quantum spins-1/2
(denoted by the set of Pauli matrices, $\{\sigma^\mu\}$) where the
Fock-space basis states are simply the classical configurations --
product states in the $\sigma^z$-basis. Assigning independent random
energies to the $2^N$ configurations leads to the random energy
model (REM)~\cite{derrida1980random} which, upon addition of spin-flip
terms $\sigma^x$ to the Hamiltonian becomes the QuantumREM
(QREM). This will be our reference unconstrained model and has no
localised phase in the bulk of the spectrum. Imposing East-like
constraints in the spin-flip terms, that is, allowing a particular
spin flip only if the spin to its right is pointing up, results in a
constrained model which we call the EastREM. The construction of the
model and a discussion of implications of the constraints for the
structure of the connectivity of Fock space constitutes
Sec.~\ref{sec:model}.

The phenomenology of the model is established in
Sec.~\ref{sec:phenomenology}. We present results for the statistics of
level spacing ratios and participation entropies of the eigenstates on
the Fock space which reveal a phase diagram with a fully localised
phase.  Dynamical autocorrelations from time evolving an initial
product state also show non-ergodic behaviour in the form of retention
of memory of initial configuration. In fact, the real-space profile of
the dynamical autocorrelation directly reflects the effect of the
corresponding local kinetic constraints. Finally, we identify the
clusters made up of densely connected states and then construct a
Hamiltonian matrix where the clusters are described by GOE
Hamiltonians but the matrix elements connecting different clusters are
as for the EastREM. This random matrix analogue to the EastREM, which
we call GOEastREM, displays the relevant features of the EastREM,
demonstrating that the clustering is the crucial ingredient.
%\al{improve this paragraph}

Analytical insights into the origin of the localisation on the Fock
space graph are obtained from the FSA, discussed in
Sec.~\ref{sec:fsa}. The FSA is an approximation for the non-local
propagator on the Fock space which takes into account the contribution
only from the shortest paths between two Fock-space sites.  As the
constraints essentially have the effect of modifying the statistics of
shortest paths on the Fock space, the FSA is ideally suited for
analysing the EastREM and exposing its differences from the QREM. As
elaborated in Sec.~\ref{sec:fsa}, two aspects of the statistics of
shortest paths are crucial, (i) the scaling of the number of Fock
space sites separated by distance $r$ with both system size and $r$,
and (ii) the scaling of number of paths between such Fock space sites
separated by $r$. These features of the Fock space are inputs to the
FSA, and the results predict an appearance of localised states in the
spectral bulk of the EastREM contrary to the QREM.  We also
corroborate the theoretical predictions from the FSA with a numerical
treatment of the FSA by enumerating the directed paths on the Fock
space, and we find that the critical point so obtained is concomitant
with that obtained from exact diagonalisation studies of
Sec.~\ref{sec:phenomenology}.

%%%%%%%%%%%%%%%%%%%%%%%%%%%%%%%%%%%%%%%%%%%%%%%%%%%%%%%%%%%%%%%%%%%%%%%%%%%%%%%%%%%%%%%%%%%%%%
%%%%%%%%%%%%%%%%%%%%%%%%%%%%%%%%%%%%%%%%%%%%%%%%%%%%%%%%%%%%%%%%%%%%%%%%%%%%%%%%%%%%%%%%%%%%%%
%%%%%%%%%%%%%%%%%%%%%%%%%%%%%%%%%%%%%%%%%%%%%%%%%%%%%%%%%%%%%%%%%%%%%%%%%%%%%%%%%%%%%%%%%%%%%%

\section{Constrained quantum model \label{sec:model}}

Our prototypical model for a kinetically constrained quantum system is
one made of $N$ spins-1/2, derived from the QREM by imposing
constraints.  The $\sigma^z$-product states constitute the basis
states of our Fock space
$\ket{\alpha}\equiv\ket{\{\sigma^z_i\}_\alpha}$ and to each of them is
associated an independent random energy $\E_\alpha$ drawn from a
normal distribution with zero mean and variance $N$.  The diagonal
(first) part of the Hamiltonian of Eq.~(\ref{eq:ham-FS}) is given by
\begin{equation}
  \hd = \sum_{\alpha=1}^{2^N}\E_\alpha\ket{\{\sigma^z_i\}_\alpha}\bra{\{\sigma^z_i\}_\alpha},
  \label{eq:hd}
\end{equation}
with $\E_\alpha\sim\mathcal{N}(0,N)$.  Henceforth, we will use the
terms spin-configuration $\ket{\{\sigma^z_i\}_\alpha}$ and Fock-space
site $\ket{\alpha}$ interchangeably.

The QREM is obtained by adding to $\hd$ unconstrained single flips
generated by the Hamiltonian
\begin{equation}
  \hunconstrained=\Gamma\sum_{i=1}^N\sigma^x_i,
  \label{eq:unconstrained-hopping}
\end{equation}
which corresponds to the second (hopping) term of
Eq.~(\ref{eq:ham-FS}), such that the total Hamiltonian is
\begin{equation}
  \hqr = \hd+\hunconstrained.
  \label{eq:qrem}
\end{equation}

In terms of Fock space sites, the QREM Hamiltonian is precisely a
$N$-dimensional hypercube with $\nh = 2^N$ vertices each of which has
a connectivity of exactly $N$: Each of the $N$ links on any vertex
corresponds to a flip of a particular spin as the single spin-flips
induced by $\hunconstrained$ are unconstrained. Another direct
implication of this is that for any Fock-space site, the number of
Fock-space sites at a distance $r$ is\footnote{Note that for the
  offdiagonal part of the Hamiltonian given by $\hunconstrained$ of
  Eq.~\eqref{eq:unconstrained-hopping}, the distance between two
  Fock-space sites is the same as the usual Hamming distance -- the
  number spins different between the two configurations.}
$\binom{N}{r}$.

Localisation or lack thereof in the QREM was studied in
Ref.~\cite{baldwin2016manybody} where it was found that the model is
ergodic in the spectral bulk for infinitesimally small $\Gamma$ while
the spectral edges can have localised eigenstates, so that there are
mobility edges at finite energy densities $\epsilon=E/N\sim
\Gamma$. However, as the width of the density of states
$\sim\sqrt{N}$, in the thermodynamic limit the localised eigenstates
occupy only a vanishing fraction of the spectrum. Generic quantum
dynamics therefore exhibit ergodic behaviour and we consider the
QREM, our reference unconstrained model to be ergodic at all
$\Gamma \neq 0$.

\begin{figure}
  \includegraphics[width=\columnwidth]{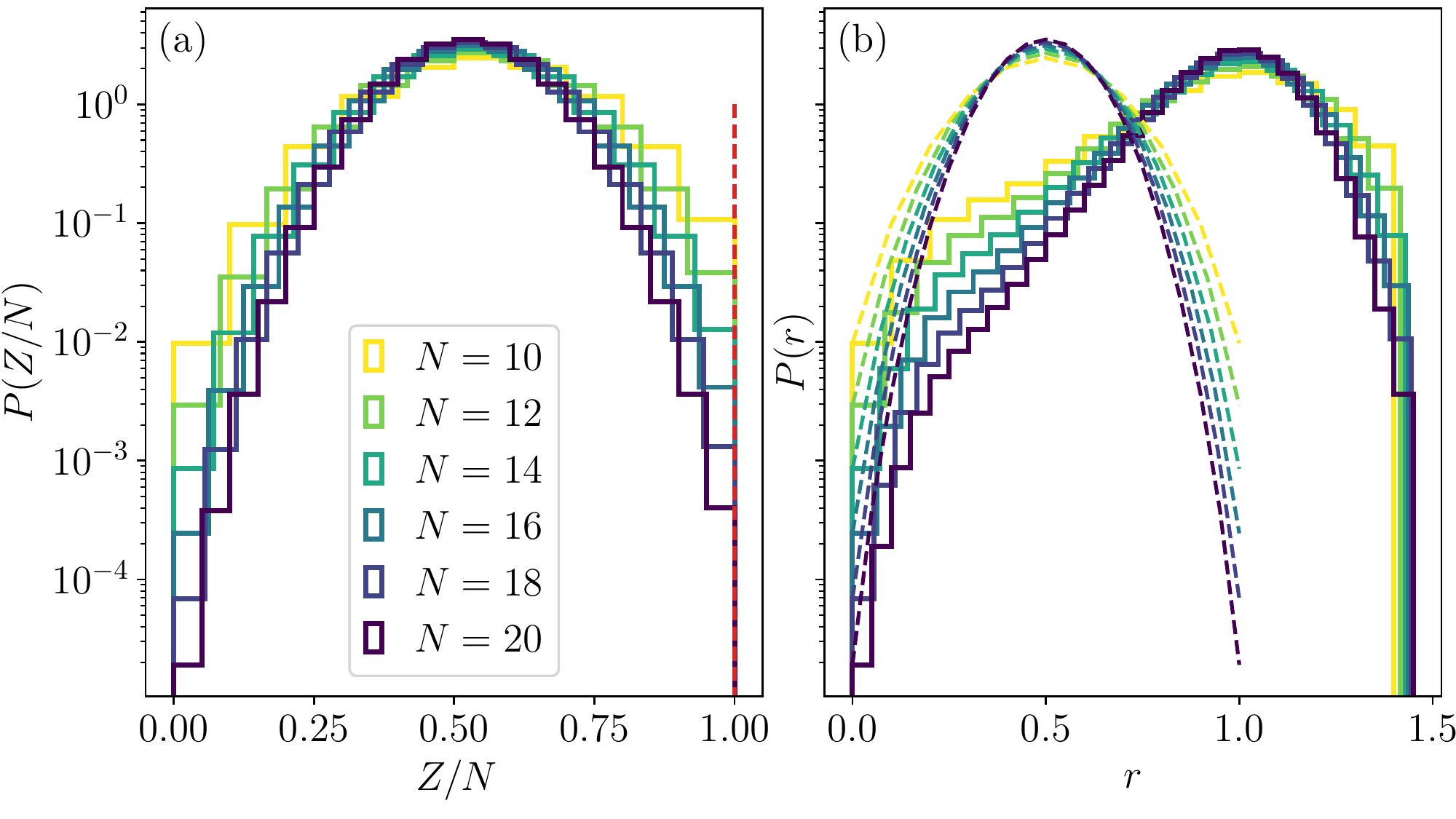}
  \caption{(a) The distribution of the connectivities $Z$ on the Fock
    space generated by $\hconstrained$ \eqref{eq:constrained-hopping}
    for different system sizes $N$. The red dashed line shows the same
    for $\hunconstrained$ \eqref{eq:unconstrained-hopping}. While for
    the latter it is a delta-function at $Z=N$, for the former it is a binomial distribution and has
    support on lower values of $Z$. (b) The distribution of the
    shortest distances from the domain wall state for $\hconstrained$
    (solid lines). The dashed lines show the same for the
    $\hunconstrained$, where it is simply the binomial distribution.}
  \label{fig:connectivity-ham-dist}
\end{figure}

The EastREM, our model for a constrained quantum system, is obtained
from the QREM by imposing local constraints of the East type such that
the Hamiltonian is given by
\begin{equation}
  \her = \hd + \hconstrained,
  \label{eq:eastrem}
\end{equation}
where
\begin{equation}
  \hconstrained=\frac{\Gamma}{2}\sum_{i=1}^N\sigma^x_i(1+\sigma^z_{i+1})
  \label{eq:constrained-hopping}
\end{equation}
where we impose periodic boundary conditions, resulting in a Fock
space that is not fragmented.\footnote{In the case of open boundary
  conditions, Fock space is fragmented as if the rightmost $w>0$ spins are all down, they remain frozen for all time.} The
constraint modifies the hopping on the Fock space (the second term in
Eq.~(\ref{eq:ham-FS})) so that it allows a spin at real-space site $i$
to be flipped if and only if the spin at site $i+1$ is pointing
up. Hence, in terms of hopping in Fock space, it has the effect of
switching off all the hopping amplitudes of the QREM Hamiltonian that
corresponded to a flip of a spin with the spin to its right pointing
up. A visual demonstration is shown in
Fig.~\ref{fig:eastrem-vs-qrem}(a) where the red links are present for
both the QREM and the EastREM while the blue links are present only on
the QREM. This has a number of consequences.

Firstly, the constraints lead to a suppression of the average
connectivity, although it still scales as $N$.  Secondly, the
distribution of connectivities, which is a delta-function at $N$ for
the QREM develops support on lower values as well for the EastREM, see
Fig.~\ref{fig:connectivity-ham-dist}(a).  In fact for the EastREM, the
distribution of connectivities is binomial
$P(Z)=\binom{N}{Z}2^{-N}$. Thirdly, the removal of the links generally
increases the shortest distance between two vertices on the Fock
space. For example, Fig.~\ref{fig:eastrem-vs-qrem}(b) shows two sites
that were a single hop away from each other on the QREM Fock-space
graph and which are much further apart on the EastREM Fock-space. This
is studied systematically in Fig.~\ref{fig:connectivity-ham-dist}(b)
where the distribution of shortest distances from a spin-configuration
has larger support on larger values for $\hconstrained$ compared to
the $\hunconstrained$.  Finally, the absence of links in the
constrained model also removes a large number of paths connecting any
two vertices (see Fig.~\ref{fig:eastrem-npath-distributions}), the
importance of which will become apparent in Sec.~\ref{sec:fsa}.  All
of the above suggest a general tendency of the constraints to localise
a state on the Fock-space. While qualitative now, these pictures will
be important later when we formalise the above ideas using the FSA on
the Fock-space.

In real space, a qualitative picture of the origins of localisation
due to the constraints is as follows. Due to the East-like
constraints, any contiguous block of down spins is slow to thermalise
as it can only do so in a sequential fashion starting from the right
edge of the block. Spins deep inside such blocks, say at a distance
$r$ away from the right edge of the block can flip only at
$r^\mathrm{th}$ order in perturbation theory. By contrast, for the
QREM any spin is free to flip and they can do so in any order.
Furthermore, even the ``liquid'' regions of the chain, which are
regions initially without such frozen blocks are affected by the
constraints dynamically. Thermalising the ``liqud'' regions involves
flipping the up spins to down creating new constrained regions, which
eventually arrest the dynamics.

\section{Phenomenology \label{sec:phenomenology}}

\subsection{Spectral properties and MBL phase
  diagram \label{subsec:spectral}}

To establish the phenomenology of the {\er} in terms of the spectral
properties and obtain an MBL phase diagram we use two commonly
studied numerical diagnostics: statistics of level spacing ratios and
participation entropies of the eigenstates on the Fock space.

The level spacing ratio, $s_n$, is defined
as~\cite{oganesyan2007localisation,pal2010many,atas2013distribution}
${s_n = \min(\Delta_n,\Delta_{n+1})/\max(\Delta_n,\Delta_{n+1})}$ with
$\Delta_n = E_n-E_{n-1}$,
where the $E_n$s denote the consecutive eigenenergies. For an ergodic
system, $s_n$ has a Wigenr-Dyson distribution, reflecting the presence
of level repulsions, so that $\braket{s}\approx 0.53$. A localised
system on the other hand has uncorrelated eigenvalues resulting in
$s_n$ having a Poisson distribution and $\braket{s}\approx0.386$.

The eigenstates on the Fock space also carry signatures of ergodicity
breaking~\cite{deluca2013ergodicity,luitz2015many,mace2019multifractal}. The
$q^\mathrm{th}$ participation entropy of an eigenstate $\ket{\psi}$
defined via
$S^\mathrm{P}_q(\ket{\psi})=\frac{1}{q-1}\ln\left[\sum_\alpha
  \vert\braket{\psi|\alpha}\vert^{2q}\right]$ scales
as~\cite{luitz2015many}
\begin{equation}
  S^\mathrm{P}_q(\ket{\psi}) = a_q\ln\nh + b_q\ln\ln\nh.
  \label{eq:pescaling}
\end{equation}
In the ergodic phase $a_q\approx 1$ as a consequence of the
eingenstate being spread over the entire Fock space whereas in the MBL
phase $a_q < 1$ indicating that the support of the eigesntate is a
vanishing fraction of the Fock space dimension in the thermodynamic
limit.

\begin{figure}
  \includegraphics[width=\columnwidth]{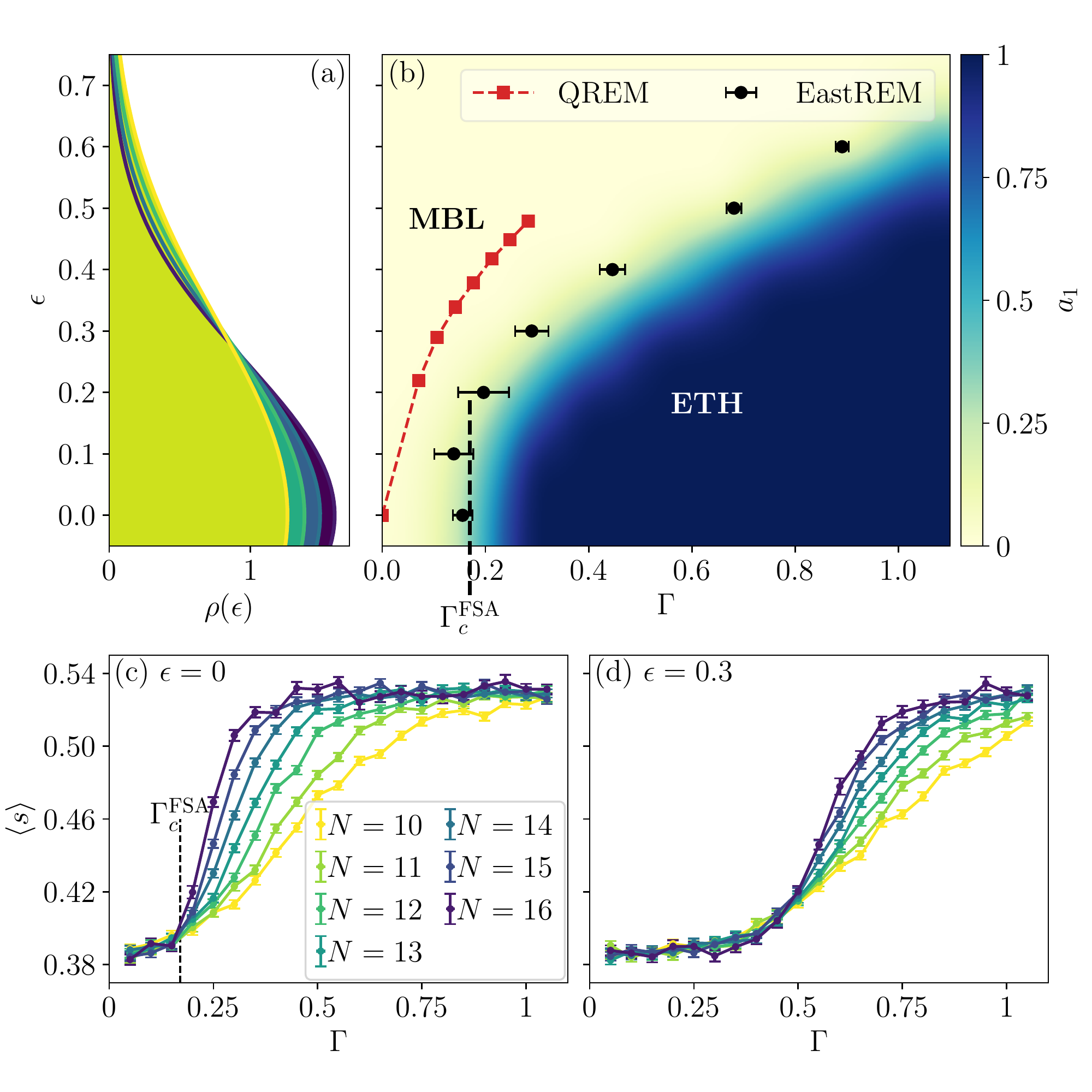}
  \caption{Localisation phase diagram of the EastREM in the
    $\Gamma$-$\epsilon$ plane. (a) The total density of states $\rho(E)$ is a Gaussian with a variance $\sim N$ ($N$ being the system size) such that in terms of energy densities $\epsilon=E/N$, the fraction of eigenstates at all finite $\epsilon$ is vanishingly small in the thermodynamic limit. Note that $\epsilon=0$ corresponds to the middle of the spectrum. (b) The ergodic region (blue) is
    characterised by the first participation ratio's volume law
    coefficient in Eq.~\eqref{eq:pescaling}, $a_1\approx 1$ as shown
    by the colour-map, whereas $a_1<1$ in the MBL phase (light
    region). The black dots show the critical $\Gamma$ extracted from
    the level spacing ratios for the EastREM whereas the red squares
    denote the critical $\Gamma$ line for the
    QREM~\cite{baldwin2017clustering}. The black dashed line denotes
    the result obtained from a numerical treatment of the FSA
    (Sec.~\ref{subsec:numerical-fsa}). (c)-(d) Representative plots of
    the mean level spacing ratio, $\braket{s}$, versus $\Gamma$ for
    different system sizes $N$ for the bulk and edges of the spectrum
    respectively. All the data was averaged over 1000 disorder
    realisations and the statistical errorbars estimated using 500
    bootstrap resamplings.}
  \label{fig:eastrem-ed}
\end{figure}

Numerically analysing the two diagnostics using exact diagonalisation
we obtain the MBL phase diagram in the $\epsilon$-$\Gamma$ plane shown
in Fig.~\ref{fig:eastrem-ed}. We emphasise that the density of states
is a Gaussian with a width proportional to $\sqrt{N}$.  Hence, any
finite energy density corresponds to the edges of the spectrum where
only a vanishing fraction of the eigenstates live in the thermodynamic
limit (see Fig.~\ref{fig:eastrem-ed}a). It is the middle of the
spectrum, $\epsilon=0$, defined via $\mathrm{Tr}[H]$, which determines the
generic dynamical behaviour of the system.

The critical $\Gamma$ can be obtained from the mean level spacing
ratio by collapsing the data for various $N$ onto a common function of
$g[(\Gamma-\Gamma_c)N^{1/\nu}]$. Such an exercise leads to the set of
critical $\Gamma_c$ at different energy densities shown by the black
circles in Fig.~\ref{fig:eastrem-ed}(b). Representative plots of the
raw data of the mean level spacing ratios in the spectral bulk and
edges are shown in panels (c) and (d) respectively.

The critical line in the $\Gamma$-$\epsilon$ plane so obtained shows a
good agreement with that of the deviation of $a_1$ from 1, the second
diagnostic for the MBL transition.  For the {\er}, a clear MBL phase
emerges at $\epsilon=0$ with a transition to the ergodic phase at
$\Gamma_c\approx 0.17$. This is qualitatively different from the QREM
where at $\epsilon=0$, the model is ergodic at all finite values of
$\Gamma$. Additionally, in the spectral edges (finite $\epsilon$), the
transition from the MBL to ergodic phase occurs at a larger value of
$\Gamma$ in the {\er} compared to the QREM; this indicates a
parametric increase of the robustness of localised phase in the
presence of the constraints.

\subsection{Non-ergodic dynamics \label{subsec:dynamics}}

\begin{figure}
  \includegraphics[width=\columnwidth]{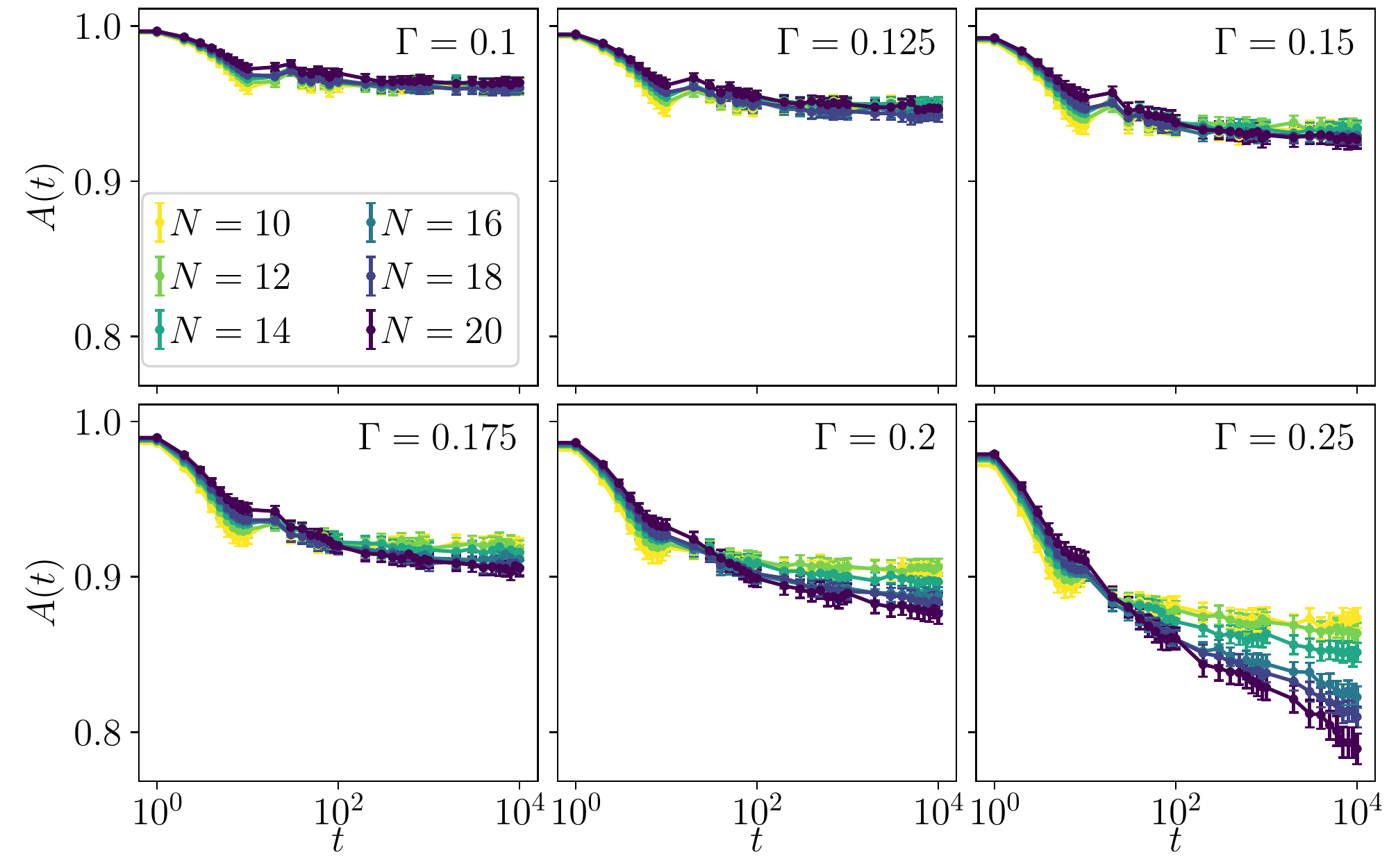}
  \caption{Autocorrelation in the EastREM as a function of time,
    $A(t)$, starting from the domain wall state for various $\Gamma$
    values and system sizes $N$. For small values of $\Gamma$, the
    $A(t)$ saturates to a finite value with no perceptible finite-size
    effects whereas for larger values of $\Gamma$, it decays with both
    $t$ and $N$ like an ergodic system. The critical $\Gamma$ can be
    estimated to lie in the vicinity of $\Gamma_c\approx 0.17$. The
    data was obtained by averaging over 1000 disorder realisations for
    all systems sizes and statistical errors estimated using standard
    bootstrap resampling.}
  \label{fig:eastrem-autocorr}
\end{figure}

As a dynamical signature of ergodicity breaking, we study the
autocorrelation function
\begin{equation}
  A(t)=\frac{1}{N}\sum_{i=1}^N\braket{\psi_0\vert\sigma^z_i(t)\sigma^z_i(0)\vert\psi_0},
  \label{eq:autocorr}
\end{equation}
where the initial state is chosen to be the domain wall (DW) state,
$\ket{\psi_0}=\ket{\underbrace{\dn\dn\cdots\dn\dn}_{N/2}\underbrace{\up\up\cdots\up\up}_{N/2}}$.
The DW state has an extensive connectivity of $N/2$ on the Fock-space
graph, so that arrested dynamics starting from this initial state, if
present, cannot be due to a subextensive connectivity of the initial
state. At the same time, it contains an extensively large blockaded
segment of down spins thus proving to be a convenient choice for
clearly demonstrating the effect of the constraints. We stress that
our choice of the initial state is not special; the phase diagram in
Fig.~\ref{fig:eastrem-ed}(a) shows that there exists a phase where
\emph{all} the eigenstates are localised.

We employ the kernel polynomial method \cite{weisse2006kernel} using
Chebyshev polynomials which allows us to evolve systems with $N=20$ up
to very long times, $t\sim 10^4$. The results for $A(t)$ are shown in
Fig.~\ref{fig:eastrem-autocorr}. For $\Gamma<\Gamma_c$, $A(t)$
saturates to a finite values at long times. The saturated value does
not depend on system size, suggesting that the system retains memory
of its initial condition in the thermodynamic limit at infinite
times. This clearly signifies a strong breaking of ergodicity.  In
contrast at larger values of $\Gamma$, $A(t)$ slowly decays with both
$t$ and $N$. The autocorrelation saturates to a finite value for
finite $N$, but this saturation value decays with $N$ such that in the
thermodynamic limit the autocorrelation decays to zero at long
times. This is the hallmark of an ergodic system. While it is
difficult to precisely determine the critical value of $\Gamma$
separating the two dynamical phases, which we estimate to be in the
vicinity of $\Gamma_c\approx 0.17$ (consistently with the exact
diagonalisation results of Sec.~\ref{subsec:spectral}), the existence
of one is clear.

\begin{figure}
  \includegraphics[width=\columnwidth]{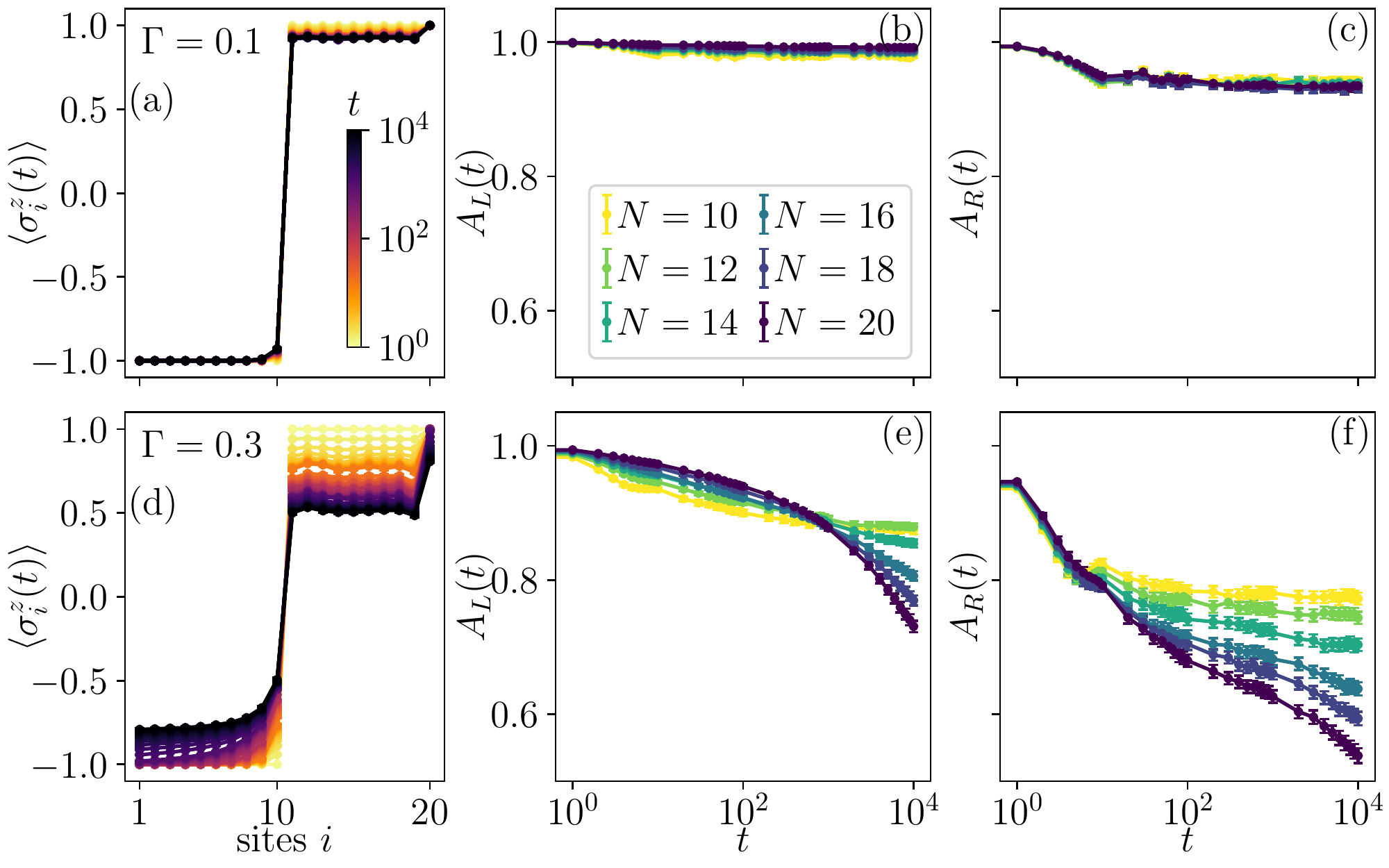}
  \caption{Panels (a) and (d) show the spin-expectation value on the
    chain (of length $N=20$) at different times (denoted by the
    colourbar) for $\Gamma=0.1$ in the localised phase and $0.3$ in
    the delocalised phase. While in the former, both the halves of the
    chain fail to thermalise, in the latter, both of them do
    thermalise although the initially blockaded left half is
    slower. The other panels show this via the behaviour of $A_L(t)$
    and $A_R(t)$, defined in Eq.~\eqref{eq:autocorr-leftright}, with
    $t$ and $N$: In the localised phase (top panels) neither half
    thermalises, while in the delocalised phase (bottom panels) both
    $L$ and $R$ eventually thermalise, with the initially blockaded
    $L$ region thermalising slower.}
  \label{fig:eastrem-autocorr-leftright}
\end{figure}

We now turn to the study of the spatially resolved dynamical
autocorrelation. We define the autocorrelation functions separately for
the left and right halves of the chain (which for the DW initial state
correspond to the blockaded and non-blockaded regions),
\begin{equation}
  A_{L(R)}(t) = \frac{2}{N}\sum_{i=1 (N/2+1)}^{N/2 (N)}\braket{\psi_0\vert\sigma^z_i(t)\sigma^z_i(0)\vert\psi_0}.
  \label{eq:autocorr-leftright}
\end{equation}
Fig.~\ref{fig:eastrem-autocorr-leftright} shows the results for
$A_{L(R)}(t)$ for two values of $\Gamma$ in both localised and
delocalised phases. In the latter both spatial regions thermalise as
reflected in their decay with $N$ and $t$, although the initially
blockaded region is much slower.  In the localised phase, both the
regions fail to thermalise as seen by the $N$-independent saturation
of both $A_L(t)$ and $A_R(t)$ at long times.

As anticipated in Sec.~\ref{sec:model}, the breaking of ergodicity
manifested in localised behaviour can be attributed to two effects.
\begin{itemize}
\item [(i)] Because of the East-like constraint of
  Eq.~\eqref{eq:constrained-hopping}, any block of contiguous down
  spins is slow to ``melt'' since the only spin in that block that can
  change dynamically is the one on the rightmost edge. The entire
  block can therefore melt only sequentially starting from the
  right. In other words, for spin-configurations with such ``solid''
  blocks of frozen spins, a large number of channels out of these
  configurations, which involve flipping of spins deep in the frozen
  block are simply unavailable. Moreover, this also has the effect of
  supressing the total number of pathways on the Fock space from one
  configuration to another. For example, there is a single shortest
  path on the Fock-space graph that connects the DW state to the
  all-up state. Contrarily for the QREM, the corresponding number of
  shortest paths is $(N/2)! \sim e^N$.
\item [(ii)] In the localised phase, the apparent liquid regions made
  up of segments of up spins also don't thermalise, see
  Fig.~\ref{fig:eastrem-autocorr-leftright}(a) and (c).  The mechanism
  underlying this is the creation of new blockades dynamically. Once a
  single spin (say at site $i$) is flipped from up to down, the one at
  $i-1$ is frozen until the $i^\mathrm{th}$ spin is flipped back
  up. However, this flipping is unavoidable; thermalising the region
  requires, by definition, that the quantum state explore all other
  spin configurations in the Fock space, and these naturally posses
  segments of down spins creating new constrained regions which
  eventually may lead to localisation.
\end{itemize}

\section{Minimally structured constrained
  model \label{subsec:clustering}}

To demonstrate the two effects mentioned at the end of
Sec.~\ref{subsec:dynamics}, we now construct a new model in which the
second effect is removed by hand while the first left in.\footnote{We
  focus on constructing a model appropriate to an initial domain wall
  state, but this is not a special choice and any spin-configuration
  could have been used.} To do so, we recognise that the first effect
above, namely the slowness of the melting of blockaded regions, is due
to the relatively small number of matrix elements leading out of
clusters of states all of which include the same blockaded island,
while the second relates to dynamics inside each cluster.

The model we construct consists of GOE matrices describing each of
these clusters, with each of these matrix blocks connected to the
others by matrix elements which are identical to the same matrix
elements as in the EastREM. It is hence a hybrid of a random matrix
with the EastREM, and arguably the least structured model that still
displays one of the features of the EastREM, namely, the difficulty of
melting the blockaded islands. Unlike the EastREM, liquid regions will
remain liquid under the dynamics of the new model, being fully chaotic
as their dynamics is described by the random matrix.

\begin{figure}
  \includegraphics[width=\columnwidth]{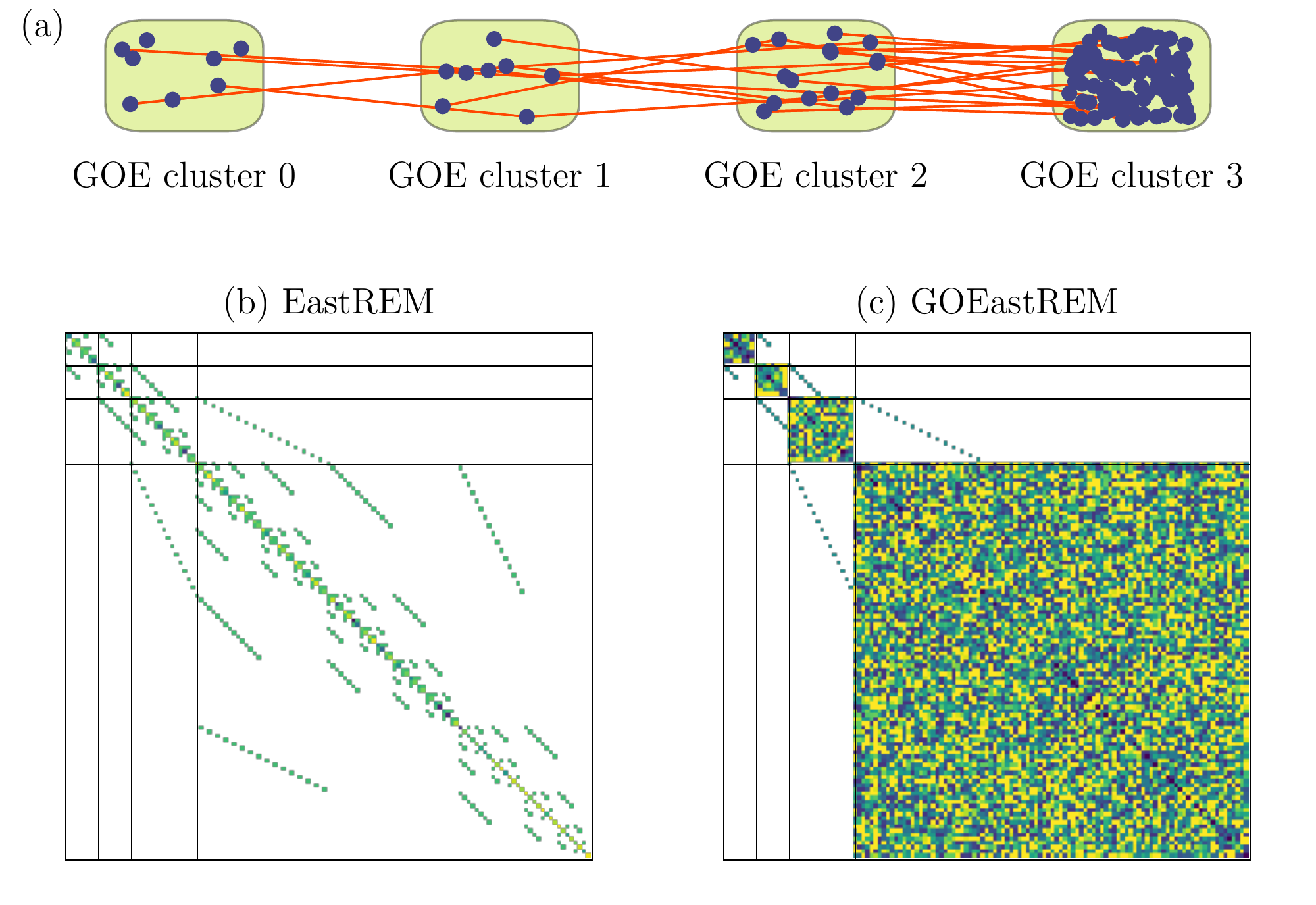}
  \caption{Construction of the GOEastREM. (a) The Fock-space graph with
    the spin-configurations arranged into clusters (denoted by the
    boxes) according to Eq.~\eqref{eq:goeastrem-cluster}. The
    offdiagonal parts of the Hamiltonian within the subspace of each
    cluster is described by a GOE matrix. The links in red denote
    hoppings on the Fock space between spin-configurations in
    different clusters as allowed by the East constraints. The
    Hamiltonian matrices of the EastREM in (b) and GOEastREM in (c) as
    a colour-map. The blocks denoted by the black lines correspond to
    the clusters in (a). Note that the inter-cluster matrix elements
    of the {\gr} are the same as that of the {\er} whereas the
    intra-cluster ones are those of GOE matrices.}
  \label{fig:goeastrem}
\end{figure}

To construct our new model, first we group together spin
configurations so that all states with a given length of blockaded
down spins starting from the leftmost spin are in the same group:
\begin{equation}
  \begin{matrix}
    {\text{Cluster\# 0}}& : & \overbrace{\dn\dn\cdots\cdots\dn\dn}^{N/2}\overbrace{\circ\circ\cdots\cdots\circ\up}^{N/2},\\
    {\text{Cluster\# 1}}& : & \overbrace{\dn\dn\cdots\cdots\dn}^{N/2-1}\up\overbrace{\circ\circ\cdots\cdots\circ\up}^{N/2},\\
    \vdots & \\
    {\text{Cluster\# $i$}}& : & \overbrace{\dn\dn\cdots\dn}^{N/2-i}\up\overbrace{\circ\circ\cdots\cdots\cdots\circ\up}^{N/2+i-1},\\
  \end{matrix}
  \label{eq:goeastrem-cluster}
\end{equation}
where the $\circ$ denotes sites which could either up or down spins.
We note two features of this separation of Fock space into
clusters:
\begin{itemize}
\item [(i)] Firstly, hoppings in the \er\ between different clusters
  correspond to progressively melting the solid block. This is because
  \er\ only allows either the rightmost spin of a blockaded island or
  the first spin after the island to flip, and either of these flips
  results in a state in cluster $i\pm 1$ so that transitions are only
  allowed between clusters $i$ and $i\pm 1$ by the \er\ rules.
\item [(ii)] Secondly, flipping spins in the liquid regions
  corresponds to Fock-space hoppings within a cluster. These lead to
  formation of new constraints as discussed in
  Sec.~\ref{subsec:dynamics} and stop the apparently liquid regions
  from thermalising in the MBL phase.
\end{itemize}

In the bottom two panels of Fig.~\ref{fig:goeastrem} we show a
representation of the Hamiltonian matrix of the \er\ (left) and \gr\
(right) in the basis of the Fock states, arranged so that states in
the same block are next to each other. The black lines correspond to
the boundaries between blocks, so that the square blocks along the
diagonal of the matrices correspond to transitions inside each
cluster while the off-diagonal blocks to transitions between the
clusters.

To allow spins to flip freely in the liquid regions without the
formation of new blockades, we randomise all matrix elements between
states in the same cluster while keeping the matrix elements between
clusters as in the \er\ model; in other words, we make the blocks on
the diagonal in Fig.~\ref{fig:goeastrem} GOE matrices while keeping
everything outside them identical to the \er. This has the effect of
allowing all intra-cluster transitions (that is, dynamics in the
liquid region) with no constraints while keeping the inter-cluster
transitions (corresponding to island melting) as in the \er\
model.\footnote{For a given disorder realisation, the diagonal
  elements, $\{\mathcal{E}_\alpha\}$, are also the same as that of the
  {\er}.} Fig.~\ref{fig:goeastrem} also makes it evident that
decreasing the size of an island by more than 1 still cannot be done
by a single application of the Hamiltonian (there are still no matrix
elements connecting clusters that are not nearest neighbours). Melting
an island is thus slow, involving a time $\mathcal{O}(\Gamma^w)$ for
an island of length $w$, like in the {\er}. On the other hand, the GOE
structure of the intra-cluster Hamiltonians means that the effect of
constraints within the cluster is no longer there as such new
constraints cannot be created in the liquid regions.

\begin{figure}
  \includegraphics[width=\columnwidth]{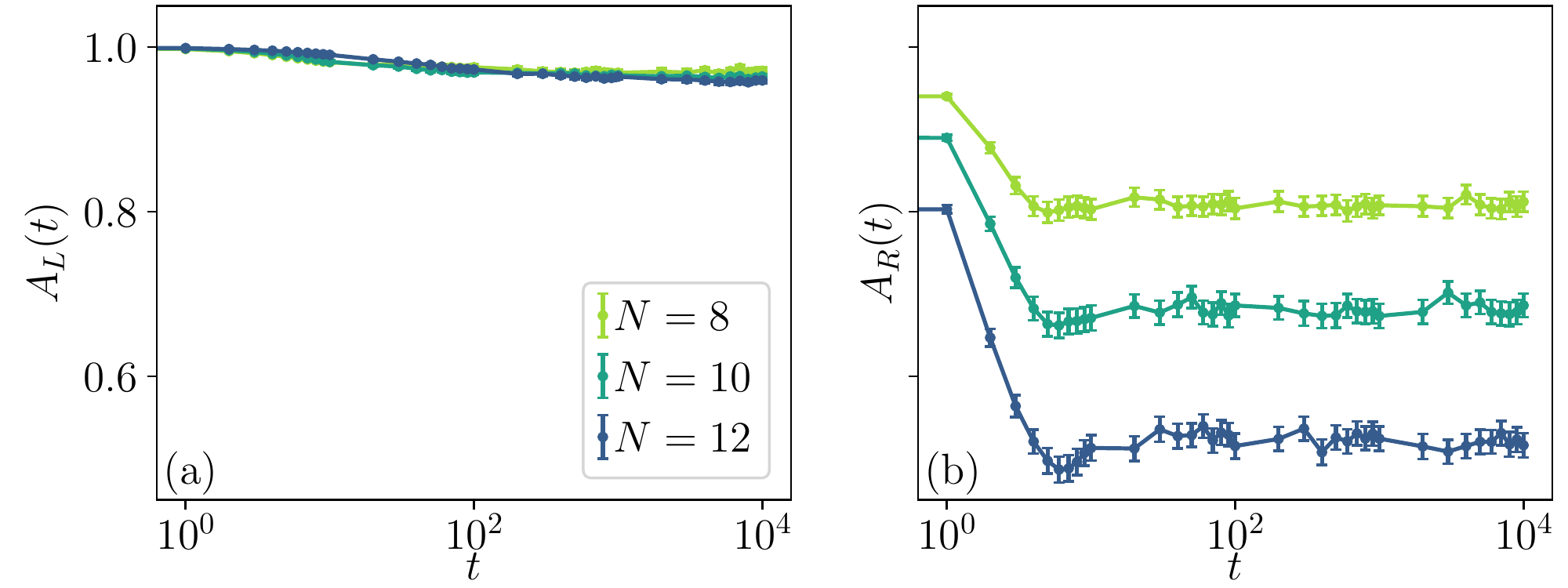}
  \caption{Dynamical autocorrelations in the left and right halves of
    the chain (Eq.~\eqref{eq:autocorr-leftright}) for the {\gr} . (a)
    The left half of the chain fails to thermalise as indicated by the
    saturation of $A_L(t)$ with both $t$ and system size $N$. (b) The
    right half rapidly thermalises as indicated by the systematic
    decay of the long-time saturation of $A_R(t)$ with $N$. Results
    shown for $\Gamma=0.1$ and data averaged over 500 disorder
    realisations.}
  \label{fig:goeastrem-autocorr-leftright}
\end{figure}

Hence out of the two effects identified earlier, namely, slow
dynamics/localisation in the already frozen region and formation
of new frozen regions, the latter has been eliminated in the {\gr}. This is
confirmed in the dynamical autocorrelations in the {\gr} starting from
the domain-wall state as shown in
Fig.~\ref{fig:goeastrem-autocorr-leftright}. The results are for
$\Gamma=0.1$ which corresponds to the MBL phase for the {\er}.  The
left half of the system which corresponds to the solid region fails to
thermalise as in the {\er} as indicated by the saturation of $A_L(t)$
with both $t$ and $N$.  On the other hand, the right half rapidly
thermalises, resulting in the systematic decay of the saturation
values of $A_R(t)$ with $N$, in stark contrast to the \er.  This
demonstrates that, as anticipated, the non-ergodic behaviour shown by
the segment of up spins in the {\er} was indeed caused by the
formation of new blockades, as the {\gr} removes that mechanism. At
the same time, as the GOEastREM preserves the constraints which lead
to non-thermalisation of segments of down spins, similar to the
EastREM, indicating that the same mechanism is at play in both the
models.

\section{Forward scattering approximation \label{sec:fsa}}

To provide analytical insight we now turn to the forward scattering
approximation (FSA), which is an approximation to the non-local (in
Fock space) Green's function to lowest order in $\Gamma$ and amounts to
a stability analysis of the trivially localised phase at vanishing
hopping $\Gamma=0$
[Eqs.~\eqref{eq:unconstrained-hopping}-\eqref{eq:eastrem}].

Considering an arbitrary initial state which we label by $\alpha=0$
and which is an eigenstate of the unperturbed $\Gamma=0$ Hamiltonian
(that is, a $\sigma^z$-product state), the weight of the perturbed
eigenstate on an arbitrary spin-configuration $\{\sigma^z_i\}_\alpha$,
denoted as $\psi(\{\sigma^z_i\}_\alpha),$ is
\begin{equation}
  \psi(\{\sigma^z_i\}_\alpha) = \sum_{p \in \text{paths}^\ast(0,\alpha)}\prod_{\beta\in p}\frac{\Gamma}{\E_0 - \E_\beta},
  \label{eq:fsa}
\end{equation}
where paths$^\ast(0,\alpha)$ is the set of all shortest paths from the
unperturbed $\alpha=0$ state to $\ket{\alpha}$. The $\E_\alpha$, as
before, are the random Fock-space site energies defined in
Eq.~\eqref{eq:hd} and are normally distributed,
$\E_\alpha\sim\mathcal{N}(0,N)$.  In this setting, the breakdown of
localisation is signalled by the probability of resonance at
arbitrarily large distances $r$ on the Fock space from the site
$\alpha=0$ approaching unity such that under the state spreads to
Fock-space sites such distances at finite $\Gamma$. The delocalisation criterion can be formally expressed as
\begin{equation}
 \lim_{r\to\infty}P\left(\frac{\ln\vert\psi_r\vert^2}{2r}>-\xi^{-1}\right)\to 1,
  \label{eq:deloc-criterion}
\end{equation}
where $\psi_r$ denotes the wavefunction amplitude on a Fock-space site
distant by $r$ from the initial state and $\xi$, an analogue of the
localisation length on the Fock space. Note that, the delocalisation
criterion of Eq.~\eqref{eq:deloc-criterion} gives a conservative
estimate in that it provides a lower bound on the critical $\Gamma$ as
it is enough for the maximum of $\psi_r$ over all configurations at
Hamming distance $r$ and disorder realisations to satisfy the
resonance condition.

\begin{figure}
  \includegraphics[width=\columnwidth]{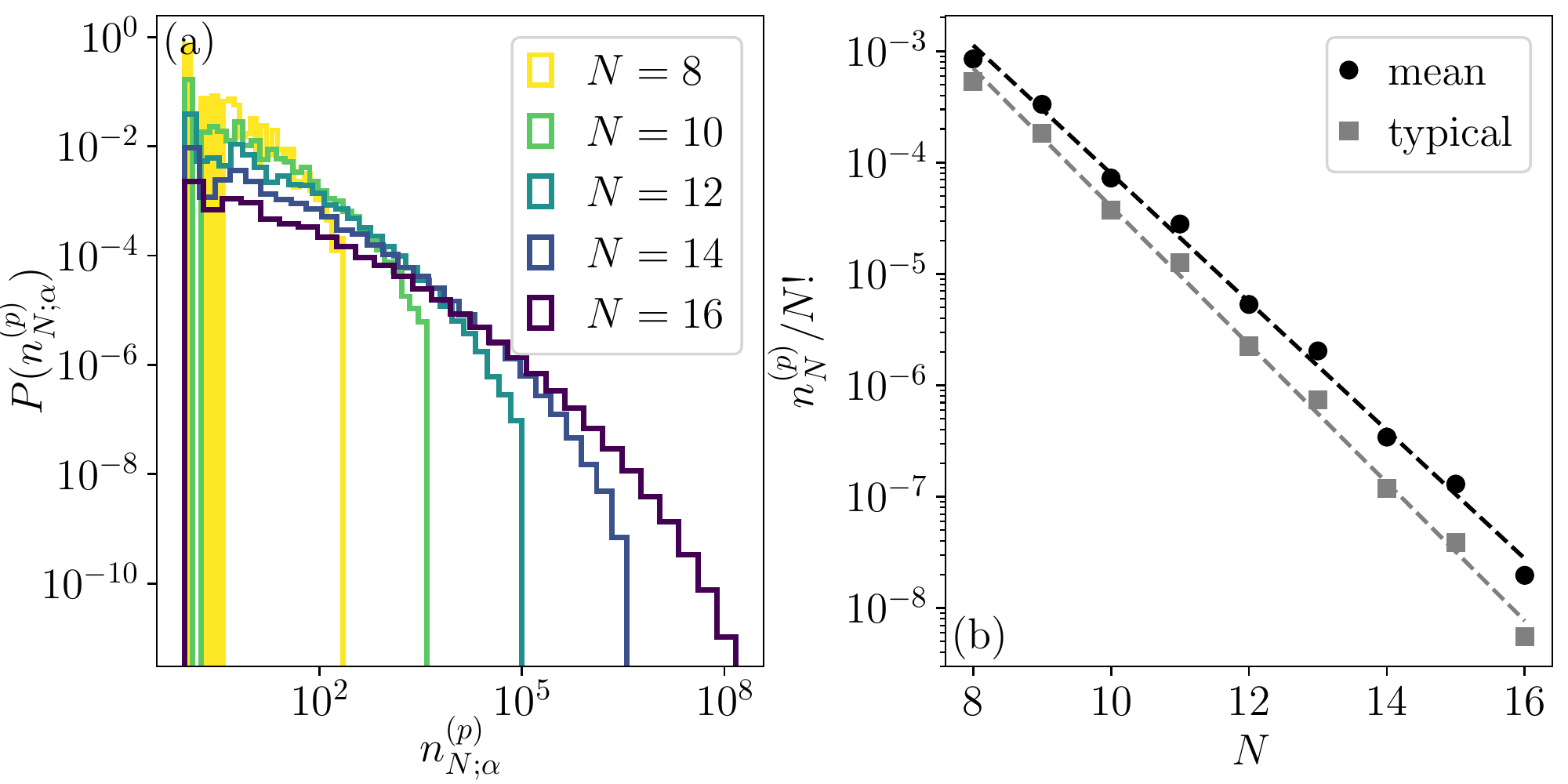}
  \caption{(a) For the EastREM, the distribution of number of paths, $\np_{r;\alpha}$, over all states $\ket{\alpha}$ at distance $r=N$ from the DW state. On a log-log scale, the curves bend downwards
    showing that the distributions decay faster than a
    power-law and hence are not fat tailed. (b) The mean and typical
    values of $\np_N$, scaled with that of the QREM ($N!$) decays
    exponentially with $N$, showing a strong suppression of the number
    of paths due to the constraints in the EastREM.}
  \label{fig:eastrem-npath-distributions}
\end{figure}

Before proceeding with the FSA analysis, it is useful to define and
assign notations to two important features of the Fock-space graph,
(i) the number of Fock-space sites at distance $r$ from the initial
state, denoted by $\ns_r$, and (ii) the number of shortest paths to
a site $\ket{\alpha}$ at distance $r$, which we denote by $\np_{r;\alpha}$. While this
quantity is different for each site, and therefore in principle
deserves its site index, Fig.~\ref{fig:eastrem-npath-distributions}
shows that its distribution is not fat tailed. We therefore omit the
site indices and use $\np_r$ to indicate the average number of paths
to sites at a distance $r$.

In the following, we analyse the localisation properties for states in
the middle of the spectrum ($\epsilon=0$) as well as the edges (finite
$\epsilon$). We find that in the middle of the spectrum (thus for the
bulk of the states and relevant regime dynamically) the {\er} has a
localisation transition at a finite $\Gamma_c$ whereas the QREM
remains delocalised for all $\Gamma$, signifying that that the
constraints change the physics qualitatively. In the edges of the
spectrum, the FSA analysis shows that localisation persists for larger
$\Gamma$ in the {\er} compared to the QREM. These two results are
consistent with those obtained from exact numerical calculations in
Sec.~\ref{sec:phenomenology}.

\subsection{Localisation by constraints in the spectral
  bulk\label{subsec:bulk}}

We first focus on states in the middle of the spectrum,
$\epsilon_0= 0$, and which constitute the majority. In this
case, all the factors $\Gamma/(\E_0-\E_\beta)\approx -\Gamma/\E_\beta$
are potentially large and individual paths can become resonant.
While a single resonant path is enough to prevent localisation in the QREM~\cite{baldwin2016manybody}, demonstrating that localisation is stable in the EastREM (Sec.~\ref{sec:phenomenology}) requires that we sum over all the paths. The probability amplitude on a state $\ket{\alpha}$ at a distance $r$ from the initial state is simply then
\begin{equation}
  \psi_\alpha = \sum_p \prod_{\beta \in p}\frac{\Gamma}{-\E_\beta},
  \label{eq:psi_alpha_edens0}
\end{equation}
where $p$ runs over all shortest paths, the lengths of which are $r$. As interference effects are not important for localisation in high dimensions, 
%the sum may be replaced by $\np_r$,
$\vert\psi_\alpha\vert = \np_r \Gamma^r
\prod_\beta(\vert\E_\beta\vert)^{-1}$. For a resonance to occur,
$\vert\psi_\alpha\vert>1$ in Eq.~\eqref{eq:psi_alpha_edens0}. Upon
defining $F_r = -\sum_{\beta=1}^r\ln\vert \E_\beta\vert$, the
resonance condition becomes $F_r>-r\ln\Gamma_r$ with
$\Gamma_r=\Gamma\left(\np_r\right)^{1/r}$. Transforming the
distributions of the independent $\E$s, the distribution of $F_r$ can
be explicitly written as
\begin{equation}
  P_F(F_r) \approx \frac{1}{(r-1)!}\left(F_r + \frac{r}{2}\ln N\right)^{r-1}e^{-\left(F_r + \frac{r}{2}\ln N\right)}.
\end{equation}
The probability for a path of length $r$ to be resonant,
$p_r^\mathrm{res}$, can be computed as the net support of the
distribution $P_F$ on $F_r\ge-r\ln\Gamma_r$,
\begin{align}
  p_r^\mathrm{res} &= \int_{-r\ln\Gamma}^{\infty}dF_r~P_F(F_r)\\
                   &\approx\frac{1}{(r-1)!}\left[r\ln\frac{\sqrt{N}}{\Gamma_r}\right]^{r-1}\exp\left[-r\ln\frac{\sqrt{N}}{\Gamma_r}\right].\label{eq:pl_edens0}
\end{align}
As each of the $\ns_r$ sites at distance $r$ are
independent, the probability that there is no resonance at distance
$r$ is given by
$Q_r=(1-p_r^\mathrm{res})^{\ns_r}\approx e^{-\ns_rp_r^\mathrm{res}}$.
The ratio
$\lambda_r = \ns_{r+1}p_{r+1}^\mathrm{res}/\ns_{r}p_{r}^\mathrm{res}$
is a monotically decreasing function of $r$. Hence, if for some $r$ we
have $\lambda_{r} <1$ then $\ns_{r}p_{r}^\mathrm{res}\to 0$ as
$r\rightarrow\infty$ and consequently $Q_r\rightarrow 1$; this signals
the stability of localisation as $Q_r$ is the probability of no
resonances at distance $r$.

Using Eq.~\eqref{eq:pl_edens0}, the localisation criterion
$\lambda_r<1$ can be rewritten as
\begin{equation}
  \Gamma \le  \underbrace{\frac{\np_r}{\np_{r+1}}\left(\frac{\ns_r}{\ns_{r+1}}\right)\frac{\sqrt{N}}{\ln N}}_{K(r,N)}\left(1+\frac{1}{r}\right)^{-r},
  \label{eq:critical-gamma-edens0}
\end{equation}
for a finite $r$ in the limit of $N\to\infty$. Hence, for localisation
to persist until a finite value of $\Gamma$, we require the dependence
of $\ns_r$ and $\np_r$ on $r$ and $N$ to be such that
$K(r,N)$ does not scale with $N$.

\begin{figure}
  \includegraphics[width=\columnwidth]{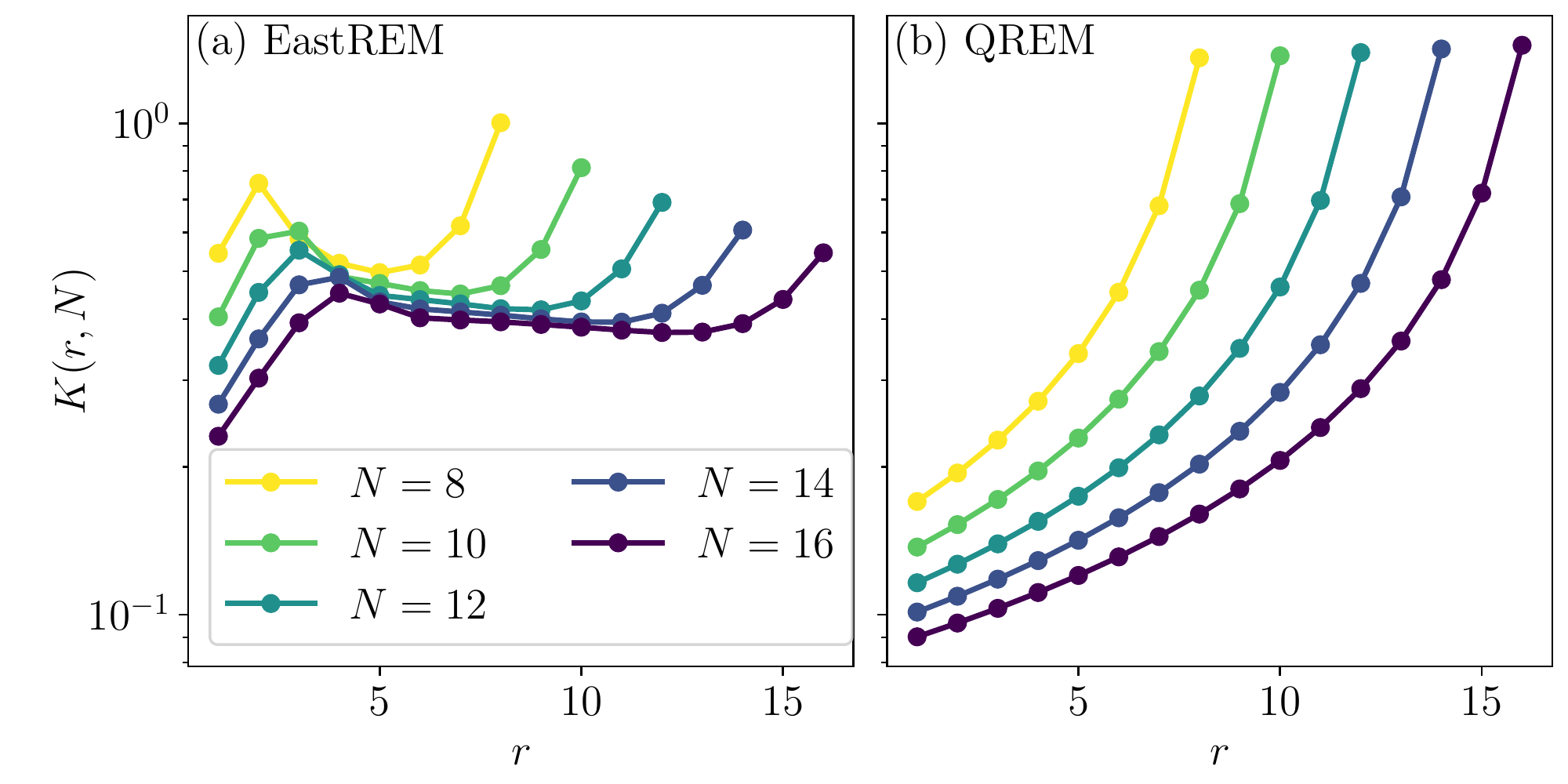}
  \caption{The factor $K(r,N)$
    appearing in Eq.~\eqref{eq:critical-gamma-edens0} for the stability of localisation. For the EastREM, there is range of finite $r$
    (seemingly growing $N$), where $K(r,N)$ does not scale with
    $N$. Contrarily in the QREM, the ratio decays with $N$ for all
    $r$.}
  \label{fig:ratio-eastrem}
\end{figure}

For the QREM, $\ns_r=\binom{N}{r}$ and $\np_r=r!$, and the ratio
$K(r,N) \sim N^{-1/2}/r$. Thus the RHS of
Eq.~\eqref{eq:critical-gamma-edens0} scales as $N^{-1/2}$ and vanishes
in the thermodynamic limit rendering localisation impossible.  On the
other hand, for the EastREM, the ratio $K(r,N)$
computed numerically does show an absence of dependence on $N$ for finite
$r$ (see Fig.~\ref{fig:ratio-eastrem}); the range of $r$ over
which this holds grows with $N$, suggesting that a
localisation-delocalisation transition is indeed possible at a finite
$\Gamma$ for $\epsilon=0$ in the thermodynamic limit.

Note that the qualitative difference between the QREM and EastREM with
regard to the ratio $\ns_r/\ns_{r+1}$ arises purely from the
constraints. In the QREM, after one flips $r\ll N$ spins, one is free
to flip any of the $N-r\approx N$ spins in the next step, which leads
to the ratio $\ns_{r+1}/\ns_r$ scaling as $N$. On the other hand, in
the EastREM, flipping some spins from up to down creates new
blockades and so the number states available on successive steps don't
scale as fast as in the QREM. This argument in conjunction with the
FSA presents an analytical picture of how the constraints affect the
distribution of distances on the Fock space which in turn lead to a
constraint-induced localised phase in the EastREM, unlike the QREM.

\subsection{Enhancement of localisation at spectral
  edges \label{subsec:edges}}

Let us now consider the situation at a finite energy density
$\epsilon_0=\E_0/N$ which, as the spectral width $\propto \sqrt{N}$,
corresponds to the edges of the spectrum. Even though the density of
states is exponentially small in  this energy region, it
is nevertheless important for the dynamics.

The distribution of $\Gamma/(\E_0-\E_\alpha)$ is fat-tailed so that
single sites can become resonant. As most of the $\E_\alpha$s are
$\sim\sqrt{N}$ these resonances are rare, so we focus on paths with a
single resonance; for a Fock-space site $\ket{\alpha}$ at distance $r$
to be resonant it is sufficient for a single site to be resonant.  For
a resonance to occur at distance $r$ (and not before), we require that
$\E_\beta\sim \sqrt{N}$ for all but the last $\beta$ on the shortest
path but $\vert\E_0-\E_\alpha\vert\ll 1$.  In this scenario, the
amplitude on the Fock-space site $\alpha$ at distance $r$ an be
expressed as
\begin{equation}
  \psi_r = \np_r\left(\frac{\Gamma}{\E_0}\right)^{r-1}\frac{\Gamma}{\E_0-\E_\alpha},
  \label{eq:psi-alpha-finite-eps}
\end{equation}
where as before $\np_r$ is the average number of paths to
sites at distance $r$ and we have implicitly assumed that all paths
are independent.  As the distribution of the number of shortest paths,
$\np_r$ is not fat-tailed (see
Fig.~\ref{fig:eastrem-npath-distributions}(a)), using the average is
justified.

From Eq.~\eqref{eq:psi-alpha-finite-eps}, a resonance at the last site requires that $\vert\psi_r\vert>1$ or
equivalently
$\vert \E_0-\E_\alpha\vert < \np_r\Gamma
\left(\frac{\Gamma}{\E_0}\right)^{r-1}$. Thus the probability of the
state being resonant is
\begin{align}
  p_r^\mathrm{res} &= \int_{\E_0-\np_r\Gamma \left(\frac{\Gamma}{\E_0}\right)^{r-1}}^{\E_0+\np_r\Gamma \left(\frac{\Gamma}{\E_0}\right)^{r-1}} d\E_\alpha \frac{1}{\sqrt{2\pi N}}\exp\left[-\frac{\E_\alpha^2}{2N}\right]\\
                   &\approx\sqrt{\frac{2}{\pi N}}\exp\left[-\frac{\E_0^2}{2N}\right]\np_r\Gamma \left(\frac{\Gamma}{\E_0}\right)^{r-1}.
                     \label{eq:pl-finite-eastrem}
\end{align}
Since the $\E$s are i.i.d. random variables, the expression above
holds for any state at a distance $r$.  The probability that none of
the $\ns_r$ sites at distance $r$ is resonant then is simply given as
before by
$Q_r = (1-p_r^\mathrm{res})^{\ns_r}\approx e^{-\ns_r
  p_r^\mathrm{res}}$.  Localisation persists if $Q_r \to 1$ as
$N \to \infty$ whenever $r$ is a finite-fraction of $N$; we thus
define $x=r/N$ which will be useful later.

\begin{figure}
  \includegraphics[width=\columnwidth]{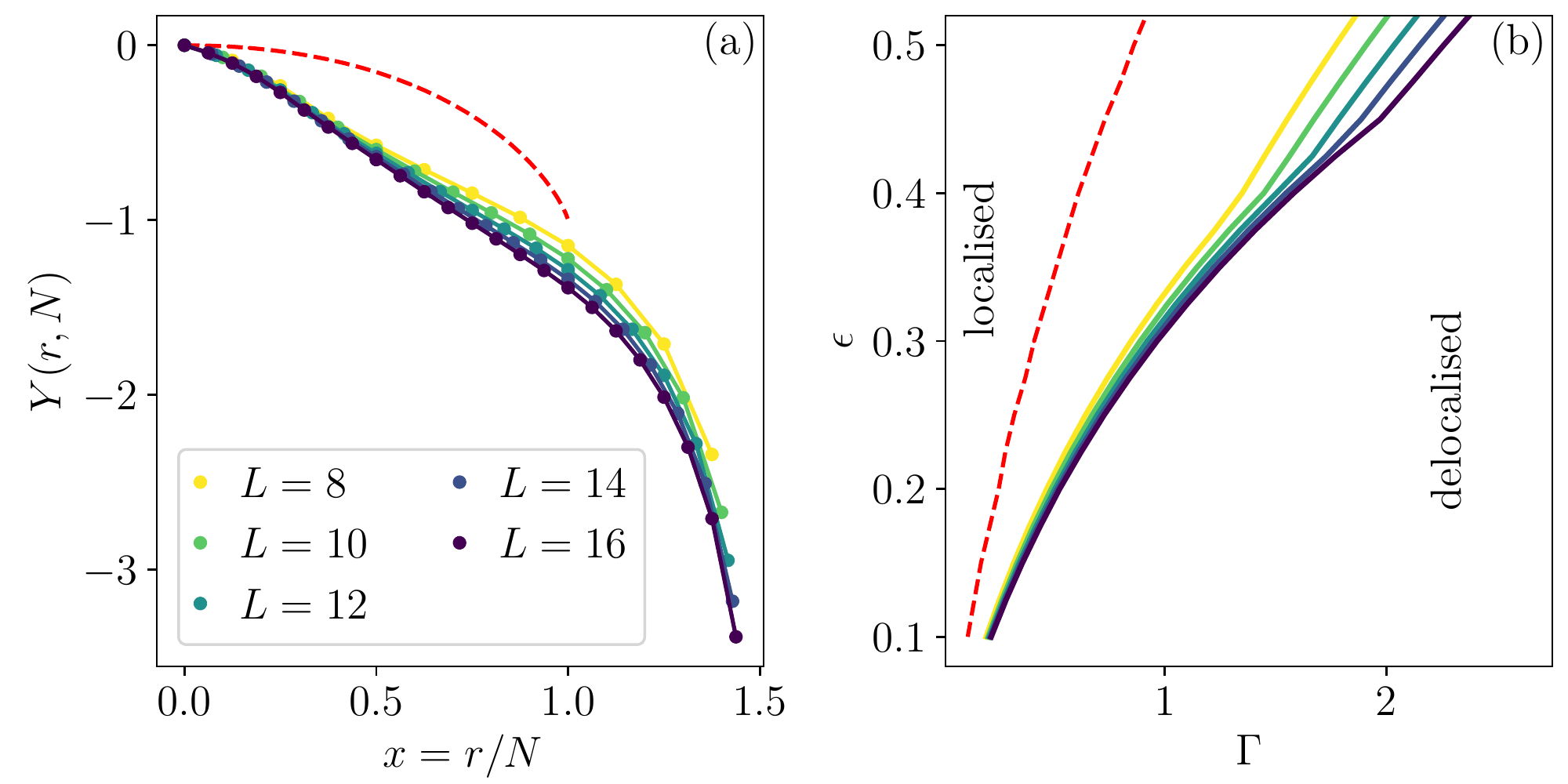}
  \caption{(a) The function $Y(r,N)$, defined in Eq.~\eqref{eq:f},
    plotted as a function of $x=r/N$ for various $N$ shows that it is
    a function of $x$ alone, for the EastREM. The red dashed line
    shows the corresponding function for the QREM. Note that $x\leq 1$
    for the QREM as no two sites are further than $N$ sites apart,
    while the removal of bonds in the \er\ allows for longer shortest
    paths. (b) The boundary between the delocalised and localised
    phases obtained from solving Eq.~\eqref{eq:gammac-trans}. As in
    (a), the red dashed line corresponds to the QREM
    result. $\Gamma_c$ for the EastREM is larger than that for the
    QREM for all finite $\epsilon_0$.}
  \label{fig:eastrem-finite-eps}
\end{figure}

Using the expression in Eq.~\eqref{eq:pl-finite-eastrem}, $Q_r$ can be
written as
\begin{equation}
  Q_r = \exp\left[-k e^{Nf(r,N,\epsilon_0)}\right],
  \label{eq:ql}
\end{equation}
where $f$ is
\begin{equation}
  f(r,\epsilon_0)=-\frac{\epsilon_0^2}{2} + x\ln\frac{\Gamma}{\epsilon_0}\underbrace{+\frac{1}{N}\ln(\ns_r \np_r)-x\ln N}_{Y(r,N)},
  \label{eq:f}
\end{equation}
where we have only kept terms that survive in the thermodynamic
limit.

Crucially, $Y(r,N)$ in Eq.~\eqref{eq:f} is only a function of $x$,
$Y(r,N)=Y(x)$, which in turn means that
$f(r,N,\epsilon_0)=f(x,\epsilon_0)$. This can be trivially shown for
the QREM using $\np_r = r!\approx (r/e)^r$ and $\ns_r = \binom{N}{r}$
that $Y(x)=-(1-x)\ln(1-x)-x$. For the EastREM such analytic expressions
$\np_r$ and $\ns_r$ are not available,\footnote{Note here that the
  mean number of shortest paths between Fock-space sites distant by
  $N$ in the EastREM scaled by the same quantity for the QREM decays
  systematically with $N$ which is a direct result of the constraints.} but the
numerically obtained form in Fig.~\ref{fig:eastrem-finite-eps}(a)
shows that $Y(r,N)$ indeed is simply a function of $x=r/N$.  The
localisation condition $Q_r\rightarrow 1$ as $N\rightarrow\infty$ then
requires that
\begin{equation}
  \max_{x}f(x,\epsilon_0)<0.
\end{equation}
The critical $\Gamma$ can be obtained by solving the equation
\begin{equation}
  -\frac{\epsilon_0^2}{2}+x_\ast\ln\frac{\Gamma_c}{\epsilon_0} + Y(x_\ast)=0
  \label{eq:gammac-trans}
\end{equation}
where $f(x,\epsilon_0)$ is maximised at $x_\ast$.

We solve Eq.~\eqref{eq:gammac-trans}, for both the QREM and the
EastREM, showing the results in Fig.~\ref{fig:eastrem-finite-eps}. We
find that in the \er, localisation persists to a larger value of
$\Gamma$.

\subsection{Numerical treatment of FSA \label{subsec:numerical-fsa}}

We now locate the transition numerically exactly within the FSA for
small system sizes. To do this, we rewrite the delocalisation
criterion of Eq.~\eqref{eq:deloc-criterion} as
\begin{equation}
  \lim_{r\to\infty}P\left(\Lambda_r > \ln\left(\frac{1}{\Gamma_c}\right) \right) \to 0,
  \label{eq:fsa-transition}
\end{equation}
where $\Lambda_r = \ln\vert\psi_r\vert^2/2r -\ln\Gamma$ and for
$\Gamma<\Gamma_c$. Our strategy is to directly calculate the
amplitudes $\psi_r$ within the FSA (by obtaining the shortest paths
numerically) and from those obtain the distribution of
Eq.~\eqref{eq:fsa-transition}. Eq.~(\ref{eq:fsa-transition}) then says
that the upper limit of its support then determines the critical
$\Gamma_c$. Without loss of generality, for this calculation we shall
take the DW state as the initial state as before.

We note here that while for the QREM to each state there corresponds
only one state at hamming distance $N$, for the EastREM there are
$\sim e^N$ such states. Indeed, in the EastREM, the number of
configurations at distance $r$ is \emph{peaked} at $r=N$. Hence, one
can argue that studying the statistics of $\Lambda_N$ can overestimate
$(1/\Gamma_c)$, thus underestimating $\Gamma_c$ as the likelihood of
having a resonance at $r=N$ is quite high simply due to a large
fraction of the configurations having $r=N$. As our main result is
that $\Gamma_c>0$ and in general larger than for the QREM,
underestimating it is not a problem.

\begin{figure}[t]
  \includegraphics[width=\columnwidth]{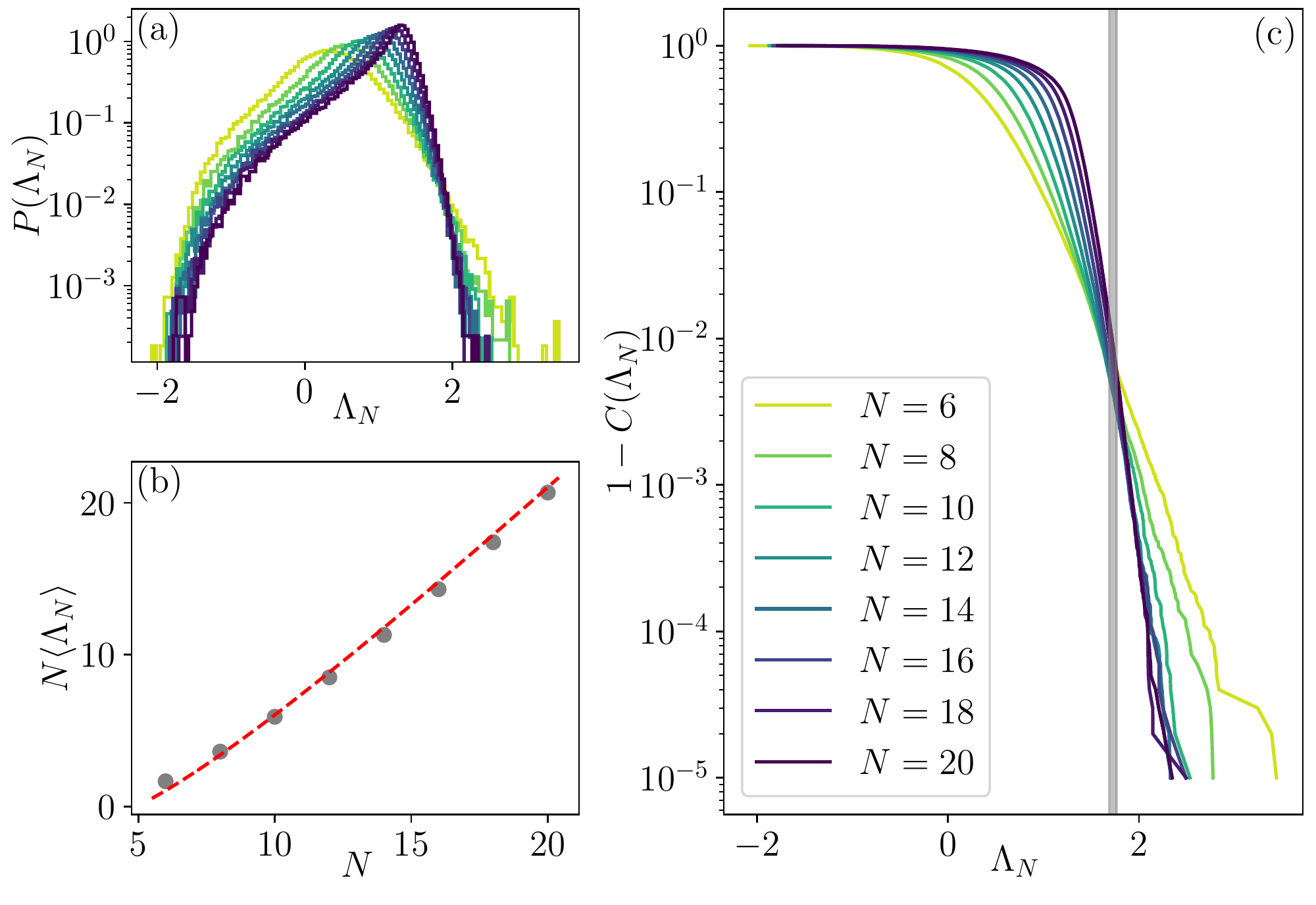}
  \caption{Numerical treatment of FSA for the EastREM. (a) The
    distributions $P(\Lambda_N)$ for different $N$. (b) The circles
    show the data for $N\braket{\Lambda_N}$ and the red dashed line
    shows a fit to the form in Eq.~\eqref{eq:fitLambdaN}. The best fit
    parameters are $\ln\left(J/\Gamma\right)_c = 1.77\pm 0.02$ and
    $\gamma=0.09$, the low value of the latter highlighting the slow
    decay of the fluctuations in $\Lambda_N$ with $N$. (c) The
    survival function, $1-C(\Lambda_N)$ corresponding to
    $P(\Lambda_N)$ shows a clear crossing of the data for various $N$
    suggesting a critical point well in agreement with that extracted
    from (b) as shown by the grey shaded region.}
  \label{fig:fsa-eastrem}
\end{figure}

To calculate $P(\Lambda_N)$ starting from the domain-wall state we
construct a matrix
\begin{equation*}
  \mathcal{T} = \Gamma\sum_{\beta,\gamma}\frac{A_{\beta,\gamma}}{\mathcal{E}_0-\mathcal{E}_\gamma}\ket{\beta}\bra{\gamma}
\end{equation*}
with $A_{\beta,\gamma}=1$ if $r_{(0,\gamma)}<r_{(0,\beta)}$ and
$\bra{\beta}\hconstrained \ket{\gamma}\neq 0$, where $r_{(0,\beta)}$
is the hamming distance between $\ket{\beta}$ and the domain-wall
state. That is, $A_{\beta,\gamma}=1$ if the transition between the two
states $\ket{\beta}$ and $\ket{\gamma}$ is allowed by the Hamiltonian
and it increases the distance from the domain-wall state.  The
amplitude on the configuration $\ket{\alpha}$, at a hamming distance
$r$ is then given by
\begin{equation*}
  \psi(\{\sigma^z_i\}_\alpha) = \bra{\alpha}\mathcal{T}^l\ket{0}; 
\end{equation*}
from which $\vert \psi_r\vert$ is obtained as
\begin{equation*}
  \vert \psi_r\vert = \max_{\alpha;~ r_{(0,\alpha)}=r}\{\psi(\{\sigma^z_i\}_\alpha)\}.
\end{equation*}
The distribution of $\Lambda_N$ so obtained is shown in
Fig.~\ref{fig:fsa-eastrem}(a) for various $N$. It clearly has finite
support, which has an upper bound that becomes sharper with increasing
$N$. This is consistent with the conclusion that as $N\to\infty$,
there exists a sharp value $1/\Gamma_c$ above which the distribution
has no weight, as required by Eq.~\eqref{eq:fsa-transition}.

Since the distribution tends to get sharper with $N$, one can argue
that the critical $1/\Gamma_c$ can be estimated as
\begin{equation}
  \lim_{N\to\infty}\braket{\Lambda_N} = \ln\left(1/\Gamma_c\right).
\end{equation}
In order to estimate this limiting value, we fit $N\braket{\Lambda_N}$
as a function of $N$ to a form
\begin{equation}
  N\braket{\Lambda_N}=a + N\ln\left(1/\Gamma\right)_c + b N^\gamma,
  \label{eq:fitLambdaN}
\end{equation}
where the last term takes into account the slowly decreasing
fluctuations in $\Lambda_N$ with increasing $N$. The fit is shown in
Fig.~\ref{fig:fsa-eastrem}(b) with the best fit parameters yielding
$\ln\left(1/\Gamma_c\right) = 1.77\pm 0.02$ which implies
$\Gamma_c = 0.17 \pm 0.01$.

Note that the transition criterion in Eq.~\eqref{eq:fsa-transition}
can also be equivalently stated as
\begin{equation}
  \lim_{N\to\infty}C(\Lambda_N = \ln(1/\Gamma)_c)\to 1,
\end{equation}
where $C(\Lambda_N)$ is the cumulative distribution corresponding to
$P(\Lambda_N)$. For finite-sized systems, we plot the $1-C(\Lambda_N)$
in Fig.~\ref{fig:fsa-eastrem}(c) and observe a clear crossing of the
data for various system sizes. The value at which the crossing occurs
and which we identify as the critical point matches remarkably well
with that obtained from the finite-size scaling analysis of
$\braket{\Lambda_N}$, as shown by the grey shaded region in
Fig.~\ref{fig:fsa-eastrem}(c).  More importantly, the critical value
so obtained, $\Gamma_c^\mathrm{FSA}$, is in excellent agreement with
the infinite temperature $\Gamma_c$ obtained from exact
diagonalisation, see Fig.~\ref{fig:eastrem-ed}.

\section{Discussion \label{sec:discussion}}

In conclusion, we have shown that local constraints can induce strong
ergodicity breaking manifested as localisation in quantum many-body
systems, crucially without shattering the Fock space.  The locality of
the constraints allows us to identify the long-lived local spatial
configurations responsible for the dynamical arrest, which in Fock
space turn out to correspond to dynamical bottlenecks caused by sparse
connectivity between clusters of states. These results are exemplified
by a quantum random energy model with East-like constraints, which we
introduce and call the {\er}.  We provide further support for the
picture of isolated regions of Fock space by constructing and studying
a random matrix model with GOE blocks for each cluster but the same
intercluster connections as the \er. This model, which we name the
{\gr}, is the minimally structured model possessing the localisation
mechanism we have identified. We finally obtain analytical insight by
applying the FSA in Fock space. The constraints modify the distance
(in Fock space) dependence of the number of accessible sites and paths
to them, and the FSA shows how this leads to localisation. In addition
to providing this insight, the FSA is in excellent agreement with the
ED results (see Fig.~\ref{fig:eastrem-ed}).

At this juncture, a number of potential directions for future work
present themselves. An immediate direction of interest is a systematic
study of the statistical mechanics of the paths on the Fock space by
treating them as directed polymers on a correlated but random
landscape. The replica trick~\cite{mezard1987spin} is ideally suited
to obtain further analytical insight into the problem as the non-local
propagator on the Fock space is expected to be dominated only by a few
paths which pass through the resonant bottlenecks.  In fact, in the
context of many-body localisation in traditionally studied
short-ranged disordered spin chains, a classical percolation proxy on
the Fock space was recently
introduced~\cite{roy2018percolation,roy2018exact}.  The effect of
constraints on such a percolation picture and the potential connections
to the directed polymer picture could shed light on the nature of the
transition.

A different question is whether an approach based on random unitary circuits, recently used to study universal properties of ergodic systems~\cite{nahum2017quantum,keyserlingk2018operator,chan2018solution}, can be generalised to include local constraints such that ergodicity is broken. In fact, there have been works on including conservation laws~\cite{rakovszky2018diffusive,khemani2018operator,friedman2019spectral} as well as ergodicity breaking~\cite{chan2018spectral} in unitary circuits. The question then is to modifying the structure of the unitary gates in the circuit such that the scrambling is constrained locally, analogously to having a conserved degree of freedom locally. The physics in this scenario remains fundamentally different from unitary circuits with conservation laws which shatter the Hilbert space~\cite{pai2019localization,khemani2019local}.

Looking further afield, periodically driven (Floquet) systems have
emerged as one of the more active areas of research in quantum
dynamics. The main difficulty in seeing interesting physics with them
is that ergodic systems inevitably heat up under
driving~\cite{lazarides2014equilibrium,dalessio2014long,ponte2015periodically}. Two routes
to arresting this heating have been
integrability~\cite{lazarides2014periodic} and
Floquet-MBL~\cite{lazarides2015fate,ponte2015many}, both of which rely on breaking ergodicity to prevent heating. It is then natural
to ask whether the present method of breaking ergodicity with local
constraints can also prevent the heating up of driven quantum systems, without explicitly fragmenting the Fock space (thus rendering the physics distinct from that of scars in Floquet systems~\cite{mukherjee2019collapse,haldar2019scars}).

Finally one might ask whether many body localisation originating from correlations in Fock space~\cite{roy2019fock}, can be identified as being caused by emergent constraints due to the correlated Fock-space disorder.

\begin{acknowledgments}
We would like to thank D.~E.~Logan for illuminating discussions about the FSA. This work was in part supported by
EPSRC Grants No. EP/N01930X/1 and EP/S020527/1.
\end{acknowledgments}

\bibliography{refs}

\begin{thebibliography}{72}%
\makeatletter
\providecommand \@ifxundefined [1]{%
 \@ifx{#1\undefined}
}%
\providecommand \@ifnum [1]{%
 \ifnum #1\expandafter \@firstoftwo
 \else \expandafter \@secondoftwo
 \fi
}%
\providecommand \@ifx [1]{%
 \ifx #1\expandafter \@firstoftwo
 \else \expandafter \@secondoftwo
 \fi
}%
\providecommand \natexlab [1]{#1}%
\providecommand \enquote  [1]{``#1''}%
\providecommand \bibnamefont  [1]{#1}%
\providecommand \bibfnamefont [1]{#1}%
\providecommand \citenamefont [1]{#1}%
\providecommand \href@noop [0]{\@secondoftwo}%
\providecommand \href [0]{\begingroup \@sanitize@url \@href}%
\providecommand \@href[1]{\@@startlink{#1}\@@href}%
\providecommand \@@href[1]{\endgroup#1\@@endlink}%
\providecommand \@sanitize@url [0]{\catcode `\\12\catcode `\$12\catcode
  `\&12\catcode `\#12\catcode `\^12\catcode `\_12\catcode `\%12\relax}%
\providecommand \@@startlink[1]{}%
\providecommand \@@endlink[0]{}%
\providecommand \url  [0]{\begingroup\@sanitize@url \@url }%
\providecommand \@url [1]{\endgroup\@href {#1}{\urlprefix }}%
\providecommand \urlprefix  [0]{URL }%
\providecommand \Eprint [0]{\href }%
\providecommand \doibase [0]{https://doi.org/}%
\providecommand \selectlanguage [0]{\@gobble}%
\providecommand \bibinfo  [0]{\@secondoftwo}%
\providecommand \bibfield  [0]{\@secondoftwo}%
\providecommand \translation [1]{[#1]}%
\providecommand \BibitemOpen [0]{}%
\providecommand \bibitemStop [0]{}%
\providecommand \bibitemNoStop [0]{.\EOS\space}%
\providecommand \EOS [0]{\spacefactor3000\relax}%
\providecommand \BibitemShut  [1]{\csname bibitem#1\endcsname}%
\let\auto@bib@innerbib\@empty
%</preamble>
\bibitem [{\citenamefont {Srednicki}(1999)}]{srednicki1999approach}%
  \BibitemOpen
  \bibfield  {author} {\bibinfo {author} {\bibfnamefont {M.}~\bibnamefont
  {Srednicki}},\ }\bibfield  {title} {\bibinfo {title} {The approach to thermal
  equilibrium in quantized chaotic systems},\ }\href
  {https://doi.org/10.1088/0305-4470/32/7/007} {\bibfield  {journal} {\bibinfo
  {journal} {J. Phys. A: Math. Gen.}\ }\textbf {\bibinfo {volume} {32}},\
  \bibinfo {pages} {1163} (\bibinfo {year} {1999})}\BibitemShut {NoStop}%
\bibitem [{\citenamefont {Rigol}\ \emph {et~al.}(2008)\citenamefont {Rigol},
  \citenamefont {Dunjko},\ and\ \citenamefont
  {Olshanii}}]{rigol2008thermalization}%
  \BibitemOpen
  \bibfield  {author} {\bibinfo {author} {\bibfnamefont {M.}~\bibnamefont
  {Rigol}}, \bibinfo {author} {\bibfnamefont {V.}~\bibnamefont {Dunjko}},\ and\
  \bibinfo {author} {\bibfnamefont {M.}~\bibnamefont {Olshanii}},\ }\bibfield
  {title} {\bibinfo {title} {Thermalization and its mechanism for generic
  isolated quantum systems},\ }\href
  {https://www.nature.com/articles/nature06838} {\bibfield  {journal} {\bibinfo
   {journal} {Nature}\ }\textbf {\bibinfo {volume} {452}},\ \bibinfo {pages}
  {854} (\bibinfo {year} {2008})}\BibitemShut {NoStop}%
\bibitem [{\citenamefont {Deutsch}(1991)}]{deutsch1991quantum}%
  \BibitemOpen
  \bibfield  {author} {\bibinfo {author} {\bibfnamefont {J.~M.}\ \bibnamefont
  {Deutsch}},\ }\bibfield  {title} {\bibinfo {title} {Quantum statistical
  mechanics in a closed system},\ }\href
  {https://doi.org/10.1103/PhysRevA.43.2046} {\bibfield  {journal} {\bibinfo
  {journal} {Phys. Rev. A}\ }\textbf {\bibinfo {volume} {43}},\ \bibinfo
  {pages} {2046} (\bibinfo {year} {1991})}\BibitemShut {NoStop}%
\bibitem [{\citenamefont {Srednicki}(1994)}]{srednicki1994chaos}%
  \BibitemOpen
  \bibfield  {author} {\bibinfo {author} {\bibfnamefont {M.}~\bibnamefont
  {Srednicki}},\ }\bibfield  {title} {\bibinfo {title} {Chaos and quantum
  thermalization},\ }\href {https://doi.org/10.1103/PhysRevE.50.888} {\bibfield
   {journal} {\bibinfo  {journal} {Phys. Rev. E}\ }\textbf {\bibinfo {volume}
  {50}},\ \bibinfo {pages} {888} (\bibinfo {year} {1994})}\BibitemShut
  {NoStop}%
\bibitem [{\citenamefont {D'Alessio}\ \emph {et~al.}(2016)\citenamefont
  {D'Alessio}, \citenamefont {Kafri}, \citenamefont {Polkovnikov},\ and\
  \citenamefont {Rigol}}]{dalessio2016quantum}%
  \BibitemOpen
  \bibfield  {author} {\bibinfo {author} {\bibfnamefont {L.}~\bibnamefont
  {D'Alessio}}, \bibinfo {author} {\bibfnamefont {Y.}~\bibnamefont {Kafri}},
  \bibinfo {author} {\bibfnamefont {A.}~\bibnamefont {Polkovnikov}},\ and\
  \bibinfo {author} {\bibfnamefont {M.}~\bibnamefont {Rigol}},\ }\bibfield
  {title} {\bibinfo {title} {From quantum chaos and eigenstate thermalization
  to statistical mechanics and thermodynamics},\ }\href
  {https://www.tandfonline.com/doi/full/10.1080/00018732.2016.1198134}
  {\bibfield  {journal} {\bibinfo  {journal} {Advances in Physics}\ }\textbf
  {\bibinfo {volume} {65}},\ \bibinfo {pages} {239} (\bibinfo {year}
  {2016})}\BibitemShut {NoStop}%
\bibitem [{\citenamefont {Deutsch}(2018)}]{deutsch2018eigenstate}%
  \BibitemOpen
  \bibfield  {author} {\bibinfo {author} {\bibfnamefont {J.~M.}\ \bibnamefont
  {Deutsch}},\ }\bibfield  {title} {\bibinfo {title} {Eigenstate thermalization
  hypothesis},\ }\href
  {http://iopscience.iop.org/article/10.1088/1361-6633/aac9f1/meta} {\bibfield
  {journal} {\bibinfo  {journal} {Rep. Prog. Phys.}\ }\textbf {\bibinfo
  {volume} {81}},\ \bibinfo {pages} {082001} (\bibinfo {year}
  {2018})}\BibitemShut {NoStop}%
\bibitem [{\citenamefont {Basko}\ \emph {et~al.}(2006)\citenamefont {Basko},
  \citenamefont {Aleiner},\ and\ \citenamefont {Altshuler}}]{basko2006metal}%
  \BibitemOpen
  \bibfield  {author} {\bibinfo {author} {\bibfnamefont {D.~M.}\ \bibnamefont
  {Basko}}, \bibinfo {author} {\bibfnamefont {I.~L.}\ \bibnamefont {Aleiner}},\
  and\ \bibinfo {author} {\bibfnamefont {B.~L.}\ \bibnamefont {Altshuler}},\
  }\bibfield  {title} {\bibinfo {title} {Metal--insulator transition in a
  weakly interacting many-electron system with localized single-particle
  states},\ }\href
  {http://www.sciencedirect.com/science/article/pii/S0003491605002630}
  {\bibfield  {journal} {\bibinfo  {journal} {Annals of {P}hysics}\ }\textbf
  {\bibinfo {volume} {321}},\ \bibinfo {pages} {1126} (\bibinfo {year}
  {2006})}\BibitemShut {NoStop}%
\bibitem [{\citenamefont {Gornyi}\ \emph {et~al.}(2005)\citenamefont {Gornyi},
  \citenamefont {Mirlin},\ and\ \citenamefont
  {Polyakov}}]{gornyi2005interacting}%
  \BibitemOpen
  \bibfield  {author} {\bibinfo {author} {\bibfnamefont {I.~V.}\ \bibnamefont
  {Gornyi}}, \bibinfo {author} {\bibfnamefont {A.~D.}\ \bibnamefont {Mirlin}},\
  and\ \bibinfo {author} {\bibfnamefont {D.~G.}\ \bibnamefont {Polyakov}},\
  }\bibfield  {title} {\bibinfo {title} {Interacting electrons in disordered
  wires: Anderson localization and low-${T}$ transport},\ }\href
  {https://doi.org/10.1103/PhysRevLett.95.206603} {\bibfield  {journal}
  {\bibinfo  {journal} {Phys. Rev. Lett.}\ }\textbf {\bibinfo {volume} {95}},\
  \bibinfo {pages} {206603} (\bibinfo {year} {2005})}\BibitemShut {NoStop}%
\bibitem [{\citenamefont {Oganesyan}\ and\ \citenamefont
  {Huse}(2007)}]{oganesyan2007localisation}%
  \BibitemOpen
  \bibfield  {author} {\bibinfo {author} {\bibfnamefont {V.}~\bibnamefont
  {Oganesyan}}\ and\ \bibinfo {author} {\bibfnamefont {D.~A.}\ \bibnamefont
  {Huse}},\ }\bibfield  {title} {\bibinfo {title} {Localization of interacting
  fermions at high temperature},\ }\href
  {https://doi.org/10.1103/PhysRevB.75.155111} {\bibfield  {journal} {\bibinfo
  {journal} {Phys. Rev. B}\ }\textbf {\bibinfo {volume} {75}},\ \bibinfo
  {pages} {155111} (\bibinfo {year} {2007})}\BibitemShut {NoStop}%
\bibitem [{\citenamefont {\ifmmode \check{Z}\else
  \v{Z}\fi{}nidari\ifmmode~\check{c}\else \v{c}\fi{}}\ \emph
  {et~al.}(2008)\citenamefont {\ifmmode \check{Z}\else
  \v{Z}\fi{}nidari\ifmmode~\check{c}\else \v{c}\fi{}}, \citenamefont {Prosen},\
  and\ \citenamefont {Prelov\ifmmode~\check{s}\else
  \v{s}\fi{}ek}}]{znidaric2008many}%
  \BibitemOpen
  \bibfield  {author} {\bibinfo {author} {\bibfnamefont {M.}~\bibnamefont
  {\ifmmode \check{Z}\else \v{Z}\fi{}nidari\ifmmode~\check{c}\else
  \v{c}\fi{}}}, \bibinfo {author} {\bibfnamefont {T.}~\bibnamefont {Prosen}},\
  and\ \bibinfo {author} {\bibfnamefont {P.}~\bibnamefont
  {Prelov\ifmmode~\check{s}\else \v{s}\fi{}ek}},\ }\bibfield  {title} {\bibinfo
  {title} {Many-body localization in the {H}eisenberg {XXZ} magnet in a random
  field},\ }\href {https://doi.org/10.1103/PhysRevB.77.064426} {\bibfield
  {journal} {\bibinfo  {journal} {Phys. Rev. B}\ }\textbf {\bibinfo {volume}
  {77}},\ \bibinfo {pages} {064426} (\bibinfo {year} {2008})}\BibitemShut
  {NoStop}%
\bibitem [{\citenamefont {Pal}\ and\ \citenamefont {Huse}(2010)}]{pal2010many}%
  \BibitemOpen
  \bibfield  {author} {\bibinfo {author} {\bibfnamefont {A.}~\bibnamefont
  {Pal}}\ and\ \bibinfo {author} {\bibfnamefont {D.~A.}\ \bibnamefont {Huse}},\
  }\bibfield  {title} {\bibinfo {title} {Many-body localization phase
  transition},\ }\href {https://doi.org/10.1103/PhysRevB.82.174411} {\bibfield
  {journal} {\bibinfo  {journal} {Phys. Rev. B}\ }\textbf {\bibinfo {volume}
  {82}},\ \bibinfo {pages} {174411} (\bibinfo {year} {2010})}\BibitemShut
  {NoStop}%
\bibitem [{\citenamefont {Nandkishore}\ and\ \citenamefont
  {Huse}(2015)}]{nandkishore2015many}%
  \BibitemOpen
  \bibfield  {author} {\bibinfo {author} {\bibfnamefont {R.}~\bibnamefont
  {Nandkishore}}\ and\ \bibinfo {author} {\bibfnamefont {D.~A.}\ \bibnamefont
  {Huse}},\ }\bibfield  {title} {\bibinfo {title} {Many-body localization and
  thermalization in quantum statistical mechanics},\ }\href
  {https://doi.org/10.1146/annurev-conmatphys-031214-014726} {\bibfield
  {journal} {\bibinfo  {journal} {Annu. Rev. Condens. Matter Phys.}\ }\textbf
  {\bibinfo {volume} {6}},\ \bibinfo {pages} {15} (\bibinfo {year}
  {2015})}\BibitemShut {NoStop}%
\bibitem [{\citenamefont {Alet}\ and\ \citenamefont
  {Laflorencie}(2018)}]{alet2018many}%
  \BibitemOpen
  \bibfield  {author} {\bibinfo {author} {\bibfnamefont {F.}~\bibnamefont
  {Alet}}\ and\ \bibinfo {author} {\bibfnamefont {N.}~\bibnamefont
  {Laflorencie}},\ }\bibfield  {title} {\bibinfo {title} {Many-body
  localization: an introduction and selected topics},\ }\href
  {https://doi.org/https://doi.org/10.1016/j.crhy.2018.03.003} {\bibfield
  {journal} {\bibinfo  {journal} {Comptes Rendus Physique}\ }\textbf {\bibinfo
  {volume} {19}},\ \bibinfo {pages} {498} (\bibinfo {year} {2018})}\BibitemShut
  {NoStop}%
\bibitem [{\citenamefont {Abanin}\ \emph {et~al.}(2019)\citenamefont {Abanin},
  \citenamefont {Altman}, \citenamefont {Bloch},\ and\ \citenamefont
  {Serbyn}}]{abanin2019colloquium}%
  \BibitemOpen
  \bibfield  {author} {\bibinfo {author} {\bibfnamefont {D.~A.}\ \bibnamefont
  {Abanin}}, \bibinfo {author} {\bibfnamefont {E.}~\bibnamefont {Altman}},
  \bibinfo {author} {\bibfnamefont {I.}~\bibnamefont {Bloch}},\ and\ \bibinfo
  {author} {\bibfnamefont {M.}~\bibnamefont {Serbyn}},\ }\bibfield  {title}
  {\bibinfo {title} {Colloquium: Many-body localization, thermalization, and
  entanglement},\ }\href {https://doi.org/10.1103/RevModPhys.91.021001}
  {\bibfield  {journal} {\bibinfo  {journal} {Rev. Mod. Phys.}\ }\textbf
  {\bibinfo {volume} {91}},\ \bibinfo {pages} {021001} (\bibinfo {year}
  {2019})}\BibitemShut {NoStop}%
\bibitem [{\citenamefont {Serbyn}\ \emph {et~al.}(2013)\citenamefont {Serbyn},
  \citenamefont {Papi\ifmmode~\acute{c}\else \'{c}\fi{}},\ and\ \citenamefont
  {Abanin}}]{serbyn2013local}%
  \BibitemOpen
  \bibfield  {author} {\bibinfo {author} {\bibfnamefont {M.}~\bibnamefont
  {Serbyn}}, \bibinfo {author} {\bibfnamefont {Z.}~\bibnamefont
  {Papi\ifmmode~\acute{c}\else \'{c}\fi{}}},\ and\ \bibinfo {author}
  {\bibfnamefont {D.~A.}\ \bibnamefont {Abanin}},\ }\bibfield  {title}
  {\bibinfo {title} {Local conservation laws and the structure of the many-body
  localized states},\ }\href {https://doi.org/10.1103/PhysRevLett.111.127201}
  {\bibfield  {journal} {\bibinfo  {journal} {Phys. Rev. Lett.}\ }\textbf
  {\bibinfo {volume} {111}},\ \bibinfo {pages} {127201} (\bibinfo {year}
  {2013})}\BibitemShut {NoStop}%
\bibitem [{\citenamefont {Huse}\ \emph {et~al.}(2014)\citenamefont {Huse},
  \citenamefont {Nandkishore},\ and\ \citenamefont
  {Oganesyan}}]{huse2014phenomenology}%
  \BibitemOpen
  \bibfield  {author} {\bibinfo {author} {\bibfnamefont {D.~A.}\ \bibnamefont
  {Huse}}, \bibinfo {author} {\bibfnamefont {R.}~\bibnamefont {Nandkishore}},\
  and\ \bibinfo {author} {\bibfnamefont {V.}~\bibnamefont {Oganesyan}},\
  }\bibfield  {title} {\bibinfo {title} {Phenomenology of fully
  many-body-localized systems},\ }\href
  {https://doi.org/10.1103/PhysRevB.90.174202} {\bibfield  {journal} {\bibinfo
  {journal} {Phys. Rev. B}\ }\textbf {\bibinfo {volume} {90}},\ \bibinfo
  {pages} {174202} (\bibinfo {year} {2014})}\BibitemShut {NoStop}%
\bibitem [{\citenamefont {Ros}\ \emph {et~al.}(2015)\citenamefont {Ros},
  \citenamefont {M{\"u}ller},\ and\ \citenamefont
  {Scardicchio}}]{ros2015integrals}%
  \BibitemOpen
  \bibfield  {author} {\bibinfo {author} {\bibfnamefont {V.}~\bibnamefont
  {Ros}}, \bibinfo {author} {\bibfnamefont {M.}~\bibnamefont {M{\"u}ller}},\
  and\ \bibinfo {author} {\bibfnamefont {A.}~\bibnamefont {Scardicchio}},\
  }\bibfield  {title} {\bibinfo {title} {Integrals of motion in the many-body
  localized phase},\ }\href
  {https://www.sciencedirect.com/science/article/pii/S0550321314003836}
  {\bibfield  {journal} {\bibinfo  {journal} {Nucl. Phys. B}\ }\textbf
  {\bibinfo {volume} {891}},\ \bibinfo {pages} {420} (\bibinfo {year}
  {2015})}\BibitemShut {NoStop}%
\bibitem [{\citenamefont {Rademaker}\ and\ \citenamefont
  {Ortu\~no}(2016)}]{rademaker2016explicit}%
  \BibitemOpen
  \bibfield  {author} {\bibinfo {author} {\bibfnamefont {L.}~\bibnamefont
  {Rademaker}}\ and\ \bibinfo {author} {\bibfnamefont {M.}~\bibnamefont
  {Ortu\~no}},\ }\bibfield  {title} {\bibinfo {title} {Explicit local integrals
  of motion for the many-body localized state},\ }\href
  {https://doi.org/10.1103/PhysRevLett.116.010404} {\bibfield  {journal}
  {\bibinfo  {journal} {Phys. Rev. Lett.}\ }\textbf {\bibinfo {volume} {116}},\
  \bibinfo {pages} {010404} (\bibinfo {year} {2016})}\BibitemShut {NoStop}%
\bibitem [{\citenamefont {Imbrie}\ \emph {et~al.}(2017)\citenamefont {Imbrie},
  \citenamefont {Ros},\ and\ \citenamefont {Scardicchio}}]{imbrie2017local}%
  \BibitemOpen
  \bibfield  {author} {\bibinfo {author} {\bibfnamefont {J.~Z.}\ \bibnamefont
  {Imbrie}}, \bibinfo {author} {\bibfnamefont {V.}~\bibnamefont {Ros}},\ and\
  \bibinfo {author} {\bibfnamefont {A.}~\bibnamefont {Scardicchio}},\
  }\bibfield  {title} {\bibinfo {title} {Local integrals of motion in many-body
  localized systems},\ }\href
  {https://onlinelibrary.wiley.com/doi/full/10.1002/andp.201600278} {\bibfield
  {journal} {\bibinfo  {journal} {Annalen der Physik}\ }\textbf {\bibinfo
  {volume} {529}},\ \bibinfo {pages} {1600278} (\bibinfo {year}
  {2017})}\BibitemShut {NoStop}%
\bibitem [{\citenamefont {Mehta}(2004)}]{mehta2004random}%
  \BibitemOpen
  \bibfield  {author} {\bibinfo {author} {\bibfnamefont {M.~L.}\ \bibnamefont
  {Mehta}},\ }\href@noop {} {\emph {\bibinfo {title} {Random matrices}}}\
  (\bibinfo  {publisher} {Elsevier},\ \bibinfo {year} {2004})\BibitemShut
  {NoStop}%
\bibitem [{\citenamefont {Logan}\ and\ \citenamefont
  {Wolynes}(1990)}]{logan1990quantum}%
  \BibitemOpen
  \bibfield  {author} {\bibinfo {author} {\bibfnamefont {D.~E.}\ \bibnamefont
  {Logan}}\ and\ \bibinfo {author} {\bibfnamefont {P.~G.}\ \bibnamefont
  {Wolynes}},\ }\bibfield  {title} {\bibinfo {title} {Quantum localization and
  energy flow in many-dimensional fermi resonant systems},\ }\href
  {https://aip.scitation.org/doi/10.1063/1.458637} {\bibfield  {journal}
  {\bibinfo  {journal} {J. Chem. Phys.}\ }\textbf {\bibinfo {volume} {93}},\
  \bibinfo {pages} {4994} (\bibinfo {year} {1990})}\BibitemShut {NoStop}%
\bibitem [{\citenamefont {Altshuler}\ \emph {et~al.}(1997)\citenamefont
  {Altshuler}, \citenamefont {Gefen}, \citenamefont {Kamenev},\ and\
  \citenamefont {Levitov}}]{altshuler1997quasiparticle}%
  \BibitemOpen
  \bibfield  {author} {\bibinfo {author} {\bibfnamefont {B.~L.}\ \bibnamefont
  {Altshuler}}, \bibinfo {author} {\bibfnamefont {Y.}~\bibnamefont {Gefen}},
  \bibinfo {author} {\bibfnamefont {A.}~\bibnamefont {Kamenev}},\ and\ \bibinfo
  {author} {\bibfnamefont {L.~S.}\ \bibnamefont {Levitov}},\ }\bibfield
  {title} {\bibinfo {title} {Quasiparticle lifetime in a finite system: A
  nonperturbative approach},\ }\href
  {https://doi.org/10.1103/PhysRevLett.78.2803} {\bibfield  {journal} {\bibinfo
   {journal} {Phys. Rev. Lett.}\ }\textbf {\bibinfo {volume} {78}},\ \bibinfo
  {pages} {2803} (\bibinfo {year} {1997})}\BibitemShut {NoStop}%
\bibitem [{\citenamefont {Logan}\ and\ \citenamefont
  {Welsh}(2019)}]{logan2019many}%
  \BibitemOpen
  \bibfield  {author} {\bibinfo {author} {\bibfnamefont {D.~E.}\ \bibnamefont
  {Logan}}\ and\ \bibinfo {author} {\bibfnamefont {S.}~\bibnamefont {Welsh}},\
  }\bibfield  {title} {\bibinfo {title} {Many-body localization in {F}ock
  space: {A} local perspective},\ }\href
  {https://doi.org/10.1103/PhysRevB.99.045131} {\bibfield  {journal} {\bibinfo
  {journal} {Phys. Rev. B}\ }\textbf {\bibinfo {volume} {99}},\ \bibinfo
  {pages} {045131} (\bibinfo {year} {2019})}\BibitemShut {NoStop}%
\bibitem [{\citenamefont {Roy}\ and\ \citenamefont
  {Logan}(2019{\natexlab{a}})}]{roy2019self}%
  \BibitemOpen
  \bibfield  {author} {\bibinfo {author} {\bibfnamefont {S.}~\bibnamefont
  {Roy}}\ and\ \bibinfo {author} {\bibfnamefont {D.~E.}\ \bibnamefont
  {Logan}},\ }\bibfield  {title} {\bibinfo {title} {{Self-consistent theory of
  many-body localisation in a quantum spin chain with long-range
  interactions}},\ }\href {https://doi.org/10.21468/SciPostPhys.7.4.042}
  {\bibfield  {journal} {\bibinfo  {journal} {SciPost Phys.}\ }\textbf
  {\bibinfo {volume} {7}},\ \bibinfo {pages} {42} (\bibinfo {year}
  {2019}{\natexlab{a}})}\BibitemShut {NoStop}%
\bibitem [{\citenamefont {Pietracaprina}\ \emph {et~al.}(2016)\citenamefont
  {Pietracaprina}, \citenamefont {Ros},\ and\ \citenamefont
  {Scardicchio}}]{pietracaprina2016forward}%
  \BibitemOpen
  \bibfield  {author} {\bibinfo {author} {\bibfnamefont {F.}~\bibnamefont
  {Pietracaprina}}, \bibinfo {author} {\bibfnamefont {V.}~\bibnamefont {Ros}},\
  and\ \bibinfo {author} {\bibfnamefont {A.}~\bibnamefont {Scardicchio}},\
  }\bibfield  {title} {\bibinfo {title} {Forward approximation as a mean-field
  approximation for the {A}nderson and many-body localization transitions},\
  }\href {https://doi.org/10.1103/PhysRevB.93.054201} {\bibfield  {journal}
  {\bibinfo  {journal} {Phys. Rev. B}\ }\textbf {\bibinfo {volume} {93}},\
  \bibinfo {pages} {054201} (\bibinfo {year} {2016})}\BibitemShut {NoStop}%
\bibitem [{\citenamefont {Pietracaprina}\ and\ \citenamefont
  {Laflorencie}(2019)}]{pietracaprina2019hilbert}%
  \BibitemOpen
  \bibfield  {author} {\bibinfo {author} {\bibfnamefont {F.}~\bibnamefont
  {Pietracaprina}}\ and\ \bibinfo {author} {\bibfnamefont {N.}~\bibnamefont
  {Laflorencie}},\ }\bibfield  {title} {\bibinfo {title} {Hilbert space
  fragmentation and many-body localization},\ }\href@noop {} {\bibfield
  {journal} {\bibinfo  {journal} {arXiv preprint arXiv:1906.05709}\ } (\bibinfo
  {year} {2019})}\BibitemShut {NoStop}%
\bibitem [{\citenamefont {Roy}\ \emph {et~al.}(2019{\natexlab{a}})\citenamefont
  {Roy}, \citenamefont {Logan},\ and\ \citenamefont {Chalker}}]{roy2018exact}%
  \BibitemOpen
  \bibfield  {author} {\bibinfo {author} {\bibfnamefont {S.}~\bibnamefont
  {Roy}}, \bibinfo {author} {\bibfnamefont {D.~E.}\ \bibnamefont {Logan}},\
  and\ \bibinfo {author} {\bibfnamefont {J.~T.}\ \bibnamefont {Chalker}},\
  }\bibfield  {title} {\bibinfo {title} {Exact solution of a percolation analog
  for the many-body localization transition},\ }\href
  {https://doi.org/10.1103/PhysRevB.99.220201} {\bibfield  {journal} {\bibinfo
  {journal} {Phys. Rev. B}\ }\textbf {\bibinfo {volume} {99}},\ \bibinfo
  {pages} {220201} (\bibinfo {year} {2019}{\natexlab{a}})}\BibitemShut
  {NoStop}%
\bibitem [{\citenamefont {Roy}\ \emph {et~al.}(2019{\natexlab{b}})\citenamefont
  {Roy}, \citenamefont {Chalker},\ and\ \citenamefont
  {Logan}}]{roy2018percolation}%
  \BibitemOpen
  \bibfield  {author} {\bibinfo {author} {\bibfnamefont {S.}~\bibnamefont
  {Roy}}, \bibinfo {author} {\bibfnamefont {J.~T.}\ \bibnamefont {Chalker}},\
  and\ \bibinfo {author} {\bibfnamefont {D.~E.}\ \bibnamefont {Logan}},\
  }\bibfield  {title} {\bibinfo {title} {Percolation in fock space as a proxy
  for many-body localization},\ }\href
  {https://doi.org/10.1103/PhysRevB.99.104206} {\bibfield  {journal} {\bibinfo
  {journal} {Phys. Rev. B}\ }\textbf {\bibinfo {volume} {99}},\ \bibinfo
  {pages} {104206} (\bibinfo {year} {2019}{\natexlab{b}})}\BibitemShut
  {NoStop}%
\bibitem [{\citenamefont {De~Tomasi}\ \emph {et~al.}(2019)\citenamefont
  {De~Tomasi}, \citenamefont {Hetterich}, \citenamefont {Sala},\ and\
  \citenamefont {Pollmann}}]{detomasi2019dynamics}%
  \BibitemOpen
  \bibfield  {author} {\bibinfo {author} {\bibfnamefont {G.}~\bibnamefont
  {De~Tomasi}}, \bibinfo {author} {\bibfnamefont {D.}~\bibnamefont
  {Hetterich}}, \bibinfo {author} {\bibfnamefont {P.}~\bibnamefont {Sala}},\
  and\ \bibinfo {author} {\bibfnamefont {F.}~\bibnamefont {Pollmann}},\
  }\bibfield  {title} {\bibinfo {title} {Dynamics of strongly interacting
  systems: From fock-space fragmentation to many-body localization},\ }\href
  {https://arxiv.org/abs/1909.03073} {\bibfield  {journal} {\bibinfo  {journal}
  {arXiv preprint arXiv:1909.03073}\ } (\bibinfo {year} {2019})}\BibitemShut
  {NoStop}%
\bibitem [{\citenamefont {Roy}\ and\ \citenamefont
  {Logan}(2019{\natexlab{b}})}]{roy2019fock}%
  \BibitemOpen
  \bibfield  {author} {\bibinfo {author} {\bibfnamefont {S.}~\bibnamefont
  {Roy}}\ and\ \bibinfo {author} {\bibfnamefont {D.~E.}\ \bibnamefont
  {Logan}},\ }\bibfield  {title} {\bibinfo {title} {Fock-space correlations and
  the origins of many-body localisation},\ }\href
  {https://arxiv.org/abs/1911.12370} {\bibfield  {journal} {\bibinfo  {journal}
  {arXiv preprint arXiv:1911.12370}\ } (\bibinfo {year}
  {2019}{\natexlab{b}})}\BibitemShut {NoStop}%
\bibitem [{\citenamefont {van Horssen}\ \emph {et~al.}(2015)\citenamefont {van
  Horssen}, \citenamefont {Levi},\ and\ \citenamefont
  {Garrahan}}]{vanhorssen2015dynamics}%
  \BibitemOpen
  \bibfield  {author} {\bibinfo {author} {\bibfnamefont {M.}~\bibnamefont {van
  Horssen}}, \bibinfo {author} {\bibfnamefont {E.}~\bibnamefont {Levi}},\ and\
  \bibinfo {author} {\bibfnamefont {J.~P.}\ \bibnamefont {Garrahan}},\
  }\bibfield  {title} {\bibinfo {title} {Dynamics of many-body localization in
  a translation-invariant quantum glass model},\ }\href
  {https://doi.org/10.1103/PhysRevB.92.100305} {\bibfield  {journal} {\bibinfo
  {journal} {Phys. Rev. B}\ }\textbf {\bibinfo {volume} {92}},\ \bibinfo
  {pages} {100305} (\bibinfo {year} {2015})}\BibitemShut {NoStop}%
\bibitem [{\citenamefont {Lan}\ \emph {et~al.}(2018)\citenamefont {Lan},
  \citenamefont {van Horssen}, \citenamefont {Powell},\ and\ \citenamefont
  {Garrahan}}]{lan2018quantum}%
  \BibitemOpen
  \bibfield  {author} {\bibinfo {author} {\bibfnamefont {Z.}~\bibnamefont
  {Lan}}, \bibinfo {author} {\bibfnamefont {M.}~\bibnamefont {van Horssen}},
  \bibinfo {author} {\bibfnamefont {S.}~\bibnamefont {Powell}},\ and\ \bibinfo
  {author} {\bibfnamefont {J.~P.}\ \bibnamefont {Garrahan}},\ }\bibfield
  {title} {\bibinfo {title} {Quantum slow relaxation and metastability due to
  dynamical constraints},\ }\href
  {https://doi.org/10.1103/PhysRevLett.121.040603} {\bibfield  {journal}
  {\bibinfo  {journal} {Phys. Rev. Lett.}\ }\textbf {\bibinfo {volume} {121}},\
  \bibinfo {pages} {040603} (\bibinfo {year} {2018})}\BibitemShut {NoStop}%
\bibitem [{\citenamefont {Pancotti}\ \emph {et~al.}(2019)\citenamefont
  {Pancotti}, \citenamefont {Giudice}, \citenamefont {Cirac}, \citenamefont
  {Garrahan},\ and\ \citenamefont {Ba{\~n}uls}}]{pancotti2019quantum}%
  \BibitemOpen
  \bibfield  {author} {\bibinfo {author} {\bibfnamefont {N.}~\bibnamefont
  {Pancotti}}, \bibinfo {author} {\bibfnamefont {G.}~\bibnamefont {Giudice}},
  \bibinfo {author} {\bibfnamefont {J.~I.}\ \bibnamefont {Cirac}}, \bibinfo
  {author} {\bibfnamefont {J.~P.}\ \bibnamefont {Garrahan}},\ and\ \bibinfo
  {author} {\bibfnamefont {M.~C.}\ \bibnamefont {Ba{\~n}uls}},\ }\bibfield
  {title} {\bibinfo {title} {Quantum east model: localization, non-thermal
  eigenstates and slow dynamics},\ }\href {https://arxiv.org/abs/1910.06616}
  {\bibfield  {journal} {\bibinfo  {journal} {arXiv preprint arXiv:1910.06616}\
  } (\bibinfo {year} {2019})}\BibitemShut {NoStop}%
\bibitem [{\citenamefont {Pai}\ \emph {et~al.}(2019)\citenamefont {Pai},
  \citenamefont {Pretko},\ and\ \citenamefont
  {Nandkishore}}]{pai2019localization}%
  \BibitemOpen
  \bibfield  {author} {\bibinfo {author} {\bibfnamefont {S.}~\bibnamefont
  {Pai}}, \bibinfo {author} {\bibfnamefont {M.}~\bibnamefont {Pretko}},\ and\
  \bibinfo {author} {\bibfnamefont {R.~M.}\ \bibnamefont {Nandkishore}},\
  }\bibfield  {title} {\bibinfo {title} {Localization in fractonic random
  circuits},\ }\href {https://doi.org/10.1103/PhysRevX.9.021003} {\bibfield
  {journal} {\bibinfo  {journal} {Phys. Rev. X}\ }\textbf {\bibinfo {volume}
  {9}},\ \bibinfo {pages} {021003} (\bibinfo {year} {2019})}\BibitemShut
  {NoStop}%
\bibitem [{\citenamefont {Khemani}\ and\ \citenamefont
  {Nandkishore}(2019)}]{khemani2019local}%
  \BibitemOpen
  \bibfield  {author} {\bibinfo {author} {\bibfnamefont {V.}~\bibnamefont
  {Khemani}}\ and\ \bibinfo {author} {\bibfnamefont {R.}~\bibnamefont
  {Nandkishore}},\ }\bibfield  {title} {\bibinfo {title} {Local constraints can
  globally shatter {H}ilbert space: a new route to quantum information
  protection},\ }\href {https://arxiv.org/abs/1904.04815} {\bibfield  {journal}
  {\bibinfo  {journal} {arXiv preprint arXiv:1904.04815}\ } (\bibinfo {year}
  {2019})}\BibitemShut {NoStop}%
\bibitem [{\citenamefont {Sala}\ \emph {et~al.}(2019)\citenamefont {Sala},
  \citenamefont {Rakovszky}, \citenamefont {Verresen}, \citenamefont {Knap},\
  and\ \citenamefont {Pollmann}}]{sala2019ergodicity}%
  \BibitemOpen
  \bibfield  {author} {\bibinfo {author} {\bibfnamefont {P.}~\bibnamefont
  {Sala}}, \bibinfo {author} {\bibfnamefont {T.}~\bibnamefont {Rakovszky}},
  \bibinfo {author} {\bibfnamefont {R.}~\bibnamefont {Verresen}}, \bibinfo
  {author} {\bibfnamefont {M.}~\bibnamefont {Knap}},\ and\ \bibinfo {author}
  {\bibfnamefont {F.}~\bibnamefont {Pollmann}},\ }\bibfield  {title} {\bibinfo
  {title} {Ergodicity-breaking arising from {H}ilbert space fragmentation in
  dipole-conserving hamiltonians},\ }\href {https://arxiv.org/abs/1904.04266}
  {\bibfield  {journal} {\bibinfo  {journal} {arXiv preprint arXiv:1904.04266}\
  } (\bibinfo {year} {2019})}\BibitemShut {NoStop}%
\bibitem [{\citenamefont {Turner}\ \emph
  {et~al.}(2018{\natexlab{a}})\citenamefont {Turner}, \citenamefont
  {Michailidis}, \citenamefont {Abanin}, \citenamefont {Serbyn},\ and\
  \citenamefont {Papi{\'c}}}]{turner2018weak}%
  \BibitemOpen
  \bibfield  {author} {\bibinfo {author} {\bibfnamefont {C.~J.}\ \bibnamefont
  {Turner}}, \bibinfo {author} {\bibfnamefont {A.~A.}\ \bibnamefont
  {Michailidis}}, \bibinfo {author} {\bibfnamefont {D.~A.}\ \bibnamefont
  {Abanin}}, \bibinfo {author} {\bibfnamefont {M.}~\bibnamefont {Serbyn}},\
  and\ \bibinfo {author} {\bibfnamefont {Z.}~\bibnamefont {Papi{\'c}}},\
  }\bibfield  {title} {\bibinfo {title} {Weak ergodicity breaking from quantum
  many-body scars},\ }\href {https://www.nature.com/articles/s41567-018-0137-5}
  {\bibfield  {journal} {\bibinfo  {journal} {Nat. Phys.}\ }\textbf {\bibinfo
  {volume} {14}},\ \bibinfo {pages} {745} (\bibinfo {year}
  {2018}{\natexlab{a}})}\BibitemShut {NoStop}%
\bibitem [{\citenamefont {Turner}\ \emph
  {et~al.}(2018{\natexlab{b}})\citenamefont {Turner}, \citenamefont
  {Michailidis}, \citenamefont {Abanin}, \citenamefont {Serbyn},\ and\
  \citenamefont {Papi\ifmmode~\acute{c}\else \'{c}\fi{}}}]{turner2018quantum}%
  \BibitemOpen
  \bibfield  {author} {\bibinfo {author} {\bibfnamefont {C.~J.}\ \bibnamefont
  {Turner}}, \bibinfo {author} {\bibfnamefont {A.~A.}\ \bibnamefont
  {Michailidis}}, \bibinfo {author} {\bibfnamefont {D.~A.}\ \bibnamefont
  {Abanin}}, \bibinfo {author} {\bibfnamefont {M.}~\bibnamefont {Serbyn}},\
  and\ \bibinfo {author} {\bibfnamefont {Z.}~\bibnamefont
  {Papi\ifmmode~\acute{c}\else \'{c}\fi{}}},\ }\bibfield  {title} {\bibinfo
  {title} {Quantum scarred eigenstates in a rydberg atom chain: Entanglement,
  breakdown of thermalization, and stability to perturbations},\ }\href
  {https://doi.org/10.1103/PhysRevB.98.155134} {\bibfield  {journal} {\bibinfo
  {journal} {Phys. Rev. B}\ }\textbf {\bibinfo {volume} {98}},\ \bibinfo
  {pages} {155134} (\bibinfo {year} {2018}{\natexlab{b}})}\BibitemShut
  {NoStop}%
\bibitem [{\citenamefont {Khemani}\ \emph {et~al.}(2019)\citenamefont
  {Khemani}, \citenamefont {Laumann},\ and\ \citenamefont
  {Chandran}}]{khemani2019signatures}%
  \BibitemOpen
  \bibfield  {author} {\bibinfo {author} {\bibfnamefont {V.}~\bibnamefont
  {Khemani}}, \bibinfo {author} {\bibfnamefont {C.~R.}\ \bibnamefont
  {Laumann}},\ and\ \bibinfo {author} {\bibfnamefont {A.}~\bibnamefont
  {Chandran}},\ }\bibfield  {title} {\bibinfo {title} {Signatures of
  integrability in the dynamics of {R}ydberg-blockaded chains},\ }\href
  {https://doi.org/10.1103/PhysRevB.99.161101} {\bibfield  {journal} {\bibinfo
  {journal} {Phys. Rev. B}\ }\textbf {\bibinfo {volume} {99}},\ \bibinfo
  {pages} {161101} (\bibinfo {year} {2019})}\BibitemShut {NoStop}%
\bibitem [{\citenamefont {Goldschmidt}(1990)}]{goldschmidt1990solvable}%
  \BibitemOpen
  \bibfield  {author} {\bibinfo {author} {\bibfnamefont {Y.~Y.}\ \bibnamefont
  {Goldschmidt}},\ }\bibfield  {title} {\bibinfo {title} {Solvable model of the
  quantum spin glass in a transverse field},\ }\href
  {https://doi.org/10.1103/PhysRevB.41.4858} {\bibfield  {journal} {\bibinfo
  {journal} {Phys. Rev. B}\ }\textbf {\bibinfo {volume} {41}},\ \bibinfo
  {pages} {4858} (\bibinfo {year} {1990})}\BibitemShut {NoStop}%
\bibitem [{\citenamefont {Laumann}\ \emph {et~al.}(2014)\citenamefont
  {Laumann}, \citenamefont {Pal},\ and\ \citenamefont
  {Scardicchio}}]{laumann2014many}%
  \BibitemOpen
  \bibfield  {author} {\bibinfo {author} {\bibfnamefont {C.~R.}\ \bibnamefont
  {Laumann}}, \bibinfo {author} {\bibfnamefont {A.}~\bibnamefont {Pal}},\ and\
  \bibinfo {author} {\bibfnamefont {A.}~\bibnamefont {Scardicchio}},\
  }\bibfield  {title} {\bibinfo {title} {Many-body mobility edge in a
  mean-field quantum spin glass},\ }\href
  {https://doi.org/10.1103/PhysRevLett.113.200405} {\bibfield  {journal}
  {\bibinfo  {journal} {Phys. Rev. Lett.}\ }\textbf {\bibinfo {volume} {113}},\
  \bibinfo {pages} {200405} (\bibinfo {year} {2014})}\BibitemShut {NoStop}%
\bibitem [{\citenamefont {Baldwin}\ \emph {et~al.}(2016)\citenamefont
  {Baldwin}, \citenamefont {Laumann}, \citenamefont {Pal},\ and\ \citenamefont
  {Scardicchio}}]{baldwin2016manybody}%
  \BibitemOpen
  \bibfield  {author} {\bibinfo {author} {\bibfnamefont {C.~L.}\ \bibnamefont
  {Baldwin}}, \bibinfo {author} {\bibfnamefont {C.~R.}\ \bibnamefont
  {Laumann}}, \bibinfo {author} {\bibfnamefont {A.}~\bibnamefont {Pal}},\ and\
  \bibinfo {author} {\bibfnamefont {A.}~\bibnamefont {Scardicchio}},\
  }\bibfield  {title} {\bibinfo {title} {The many-body localized phase of the
  quantum random energy model},\ }\href
  {https://doi.org/10.1103/PhysRevB.93.024202} {\bibfield  {journal} {\bibinfo
  {journal} {Phys. Rev. B}\ }\textbf {\bibinfo {volume} {93}},\ \bibinfo
  {pages} {024202} (\bibinfo {year} {2016})}\BibitemShut {NoStop}%
\bibitem [{\citenamefont {Baldwin}\ \emph {et~al.}(2017)\citenamefont
  {Baldwin}, \citenamefont {Laumann}, \citenamefont {Pal},\ and\ \citenamefont
  {Scardicchio}}]{baldwin2017clustering}%
  \BibitemOpen
  \bibfield  {author} {\bibinfo {author} {\bibfnamefont {C.~L.}\ \bibnamefont
  {Baldwin}}, \bibinfo {author} {\bibfnamefont {C.~R.}\ \bibnamefont
  {Laumann}}, \bibinfo {author} {\bibfnamefont {A.}~\bibnamefont {Pal}},\ and\
  \bibinfo {author} {\bibfnamefont {A.}~\bibnamefont {Scardicchio}},\
  }\bibfield  {title} {\bibinfo {title} {Clustering of nonergodic eigenstates
  in quantum spin glasses},\ }\href
  {https://doi.org/10.1103/PhysRevLett.118.127201} {\bibfield  {journal}
  {\bibinfo  {journal} {Phys. Rev. Lett.}\ }\textbf {\bibinfo {volume} {118}},\
  \bibinfo {pages} {127201} (\bibinfo {year} {2017})}\BibitemShut {NoStop}%
\bibitem [{\citenamefont {Ritort}\ and\ \citenamefont
  {Sollich}(2003)}]{ritort2003glassy}%
  \BibitemOpen
  \bibfield  {author} {\bibinfo {author} {\bibfnamefont {F.}~\bibnamefont
  {Ritort}}\ and\ \bibinfo {author} {\bibfnamefont {P.}~\bibnamefont
  {Sollich}},\ }\bibfield  {title} {\bibinfo {title} {Glassy dynamics of
  kinetically constrained models},\ }\href
  {https://www.tandfonline.com/doi/abs/10.1080/0001873031000093582} {\bibfield
  {journal} {\bibinfo  {journal} {Advances in Physics}\ }\textbf {\bibinfo
  {volume} {52}},\ \bibinfo {pages} {219} (\bibinfo {year} {2003})}\BibitemShut
  {NoStop}%
\bibitem [{\citenamefont {Garrahan}\ \emph {et~al.}(2011)\citenamefont
  {Garrahan}, \citenamefont {Sollich},\ and\ \citenamefont
  {Toninelli}}]{garrahan2011kinetically}%
  \BibitemOpen
  \bibfield  {author} {\bibinfo {author} {\bibfnamefont {J.~P.}\ \bibnamefont
  {Garrahan}}, \bibinfo {author} {\bibfnamefont {P.}~\bibnamefont {Sollich}},\
  and\ \bibinfo {author} {\bibfnamefont {C.}~\bibnamefont {Toninelli}},\
  }\bibfield  {title} {\bibinfo {title} {Kinetically constrained models},\ }in\
  \href
  {http://www.oxfordscholarship.com/view/10.1093/acprof:oso/9780199691470.001.0001/acprof-9780199691470-chapter-10}
  {\emph {\bibinfo {booktitle} {Dynamical heterogeneities in glasses, colloids,
  and granular media}}},\ \bibinfo {editor} {edited by\ \bibinfo {editor}
  {\bibfnamefont {L.}~\bibnamefont {Berthier}}, \bibinfo {editor}
  {\bibfnamefont {G.}~\bibnamefont {Biroli}}, \bibinfo {editor} {\bibfnamefont
  {J.-P.}\ \bibnamefont {Bouchaud}}, \bibinfo {editor} {\bibfnamefont
  {L.}~\bibnamefont {Cipelletti}},\ and\ \bibinfo {editor} {\bibfnamefont
  {W.}~\bibnamefont {van Saarloos}}}\ (\bibinfo  {publisher} {Oxford University
  Press},\ \bibinfo {address} {Oxford},\ \bibinfo {year} {2011})\BibitemShut
  {NoStop}%
\bibitem [{\citenamefont {Garrahan}(2018)}]{garrahan2018aspects}%
  \BibitemOpen
  \bibfield  {author} {\bibinfo {author} {\bibfnamefont {J.~P.}\ \bibnamefont
  {Garrahan}},\ }\bibfield  {title} {\bibinfo {title} {Aspects of
  non-equilibrium in classical and quantum systems: Slow relaxation and
  glasses, dynamical large deviations, quantum non-ergodicity, and open quantum
  dynamics},\ }\href
  {https://www.sciencedirect.com/science/article/pii/S0378437117313985}
  {\bibfield  {journal} {\bibinfo  {journal} {Physica A: Statistical Mechanics
  and its Applications}\ }\textbf {\bibinfo {volume} {504}},\ \bibinfo {pages}
  {130} (\bibinfo {year} {2018})}\BibitemShut {NoStop}%
\bibitem [{\citenamefont {Abou-Chacra}\ \emph {et~al.}(1973)\citenamefont
  {Abou-Chacra}, \citenamefont {Thouless},\ and\ \citenamefont
  {Anderson}}]{abou-chacra1973self}%
  \BibitemOpen
  \bibfield  {author} {\bibinfo {author} {\bibfnamefont {R.}~\bibnamefont
  {Abou-Chacra}}, \bibinfo {author} {\bibfnamefont {D.~J.}\ \bibnamefont
  {Thouless}},\ and\ \bibinfo {author} {\bibfnamefont {P.~W.}\ \bibnamefont
  {Anderson}},\ }\bibfield  {title} {\bibinfo {title} {A self-consistent theory
  of localization},\ }\href {https://doi.org/10.1088/0022-3719/6/10/009}
  {\bibfield  {journal} {\bibinfo  {journal} {Journal of Physics C: Solid State
  Physics}\ }\textbf {\bibinfo {volume} {6}},\ \bibinfo {pages} {1734}
  (\bibinfo {year} {1973})}\BibitemShut {NoStop}%
\bibitem [{\citenamefont {Anderson}(1958)}]{anderson1958absence}%
  \BibitemOpen
  \bibfield  {author} {\bibinfo {author} {\bibfnamefont {P.~W.}\ \bibnamefont
  {Anderson}},\ }\bibfield  {title} {\bibinfo {title} {Absence of diffusion in
  certain random lattices},\ }\href {https://doi.org/10.1103/PhysRev.109.1492}
  {\bibfield  {journal} {\bibinfo  {journal} {Phys. Rev.}\ }\textbf {\bibinfo
  {volume} {109}},\ \bibinfo {pages} {1492} (\bibinfo {year}
  {1958})}\BibitemShut {NoStop}%
\bibitem [{\citenamefont {Luitz}\ \emph {et~al.}(2015)\citenamefont {Luitz},
  \citenamefont {Laflorencie},\ and\ \citenamefont {Alet}}]{luitz2015many}%
  \BibitemOpen
  \bibfield  {author} {\bibinfo {author} {\bibfnamefont {D.~J.}\ \bibnamefont
  {Luitz}}, \bibinfo {author} {\bibfnamefont {N.}~\bibnamefont {Laflorencie}},\
  and\ \bibinfo {author} {\bibfnamefont {F.}~\bibnamefont {Alet}},\ }\bibfield
  {title} {\bibinfo {title} {Many-body localization edge in the random-field
  {H}eisenberg chain},\ }\href {https://doi.org/10.1103/PhysRevB.91.081103}
  {\bibfield  {journal} {\bibinfo  {journal} {Phys. Rev. B}\ }\textbf {\bibinfo
  {volume} {91}},\ \bibinfo {pages} {081103} (\bibinfo {year}
  {2015})}\BibitemShut {NoStop}%
\bibitem [{\citenamefont {Bar~Lev}\ \emph {et~al.}(2015)\citenamefont
  {Bar~Lev}, \citenamefont {Cohen},\ and\ \citenamefont
  {Reichman}}]{lev2015absence}%
  \BibitemOpen
  \bibfield  {author} {\bibinfo {author} {\bibfnamefont {Y.}~\bibnamefont
  {Bar~Lev}}, \bibinfo {author} {\bibfnamefont {G.}~\bibnamefont {Cohen}},\
  and\ \bibinfo {author} {\bibfnamefont {D.~R.}\ \bibnamefont {Reichman}},\
  }\bibfield  {title} {\bibinfo {title} {Absence of diffusion in an interacting
  system of spinless fermions on a one-dimensional disordered lattice},\ }\href
  {https://doi.org/10.1103/PhysRevLett.114.100601} {\bibfield  {journal}
  {\bibinfo  {journal} {Phys. Rev. Lett.}\ }\textbf {\bibinfo {volume} {114}},\
  \bibinfo {pages} {100601} (\bibinfo {year} {2015})}\BibitemShut {NoStop}%
\bibitem [{\citenamefont {Imbrie}(2016)}]{imbrie2016many}%
  \BibitemOpen
  \bibfield  {author} {\bibinfo {author} {\bibfnamefont {J.~Z.}\ \bibnamefont
  {Imbrie}},\ }\bibfield  {title} {\bibinfo {title} {On many-body localization
  for quantum spin chains},\ }\href
  {https://link.springer.com/article/10.1007%2Fs10955-016-1508-x} {\bibfield
  {journal} {\bibinfo  {journal} {J. Stat. Phys.}\ }\textbf {\bibinfo {volume}
  {163}},\ \bibinfo {pages} {998} (\bibinfo {year} {2016})}\BibitemShut
  {NoStop}%
\bibitem [{\citenamefont {Derrida}(1980)}]{derrida1980random}%
  \BibitemOpen
  \bibfield  {author} {\bibinfo {author} {\bibfnamefont {B.}~\bibnamefont
  {Derrida}},\ }\bibfield  {title} {\bibinfo {title} {Random-energy model:
  Limit of a family of disordered models},\ }\href
  {https://doi.org/10.1103/PhysRevLett.45.79} {\bibfield  {journal} {\bibinfo
  {journal} {Phys. Rev. Lett.}\ }\textbf {\bibinfo {volume} {45}},\ \bibinfo
  {pages} {79} (\bibinfo {year} {1980})}\BibitemShut {NoStop}%
\bibitem [{\citenamefont {Atas}\ \emph {et~al.}(2013)\citenamefont {Atas},
  \citenamefont {Bogomolny}, \citenamefont {Giraud},\ and\ \citenamefont
  {Roux}}]{atas2013distribution}%
  \BibitemOpen
  \bibfield  {author} {\bibinfo {author} {\bibfnamefont {Y.~Y.}\ \bibnamefont
  {Atas}}, \bibinfo {author} {\bibfnamefont {E.}~\bibnamefont {Bogomolny}},
  \bibinfo {author} {\bibfnamefont {O.}~\bibnamefont {Giraud}},\ and\ \bibinfo
  {author} {\bibfnamefont {G.}~\bibnamefont {Roux}},\ }\bibfield  {title}
  {\bibinfo {title} {Distribution of the ratio of consecutive level spacings in
  random matrix ensembles},\ }\href
  {https://doi.org/10.1103/PhysRevLett.110.084101} {\bibfield  {journal}
  {\bibinfo  {journal} {Phys. Rev. Lett.}\ }\textbf {\bibinfo {volume} {110}},\
  \bibinfo {pages} {084101} (\bibinfo {year} {2013})}\BibitemShut {NoStop}%
\bibitem [{\citenamefont {De~Luca}\ and\ \citenamefont
  {Scardicchio}(2013)}]{deluca2013ergodicity}%
  \BibitemOpen
  \bibfield  {author} {\bibinfo {author} {\bibfnamefont {A.}~\bibnamefont
  {De~Luca}}\ and\ \bibinfo {author} {\bibfnamefont {A.}~\bibnamefont
  {Scardicchio}},\ }\bibfield  {title} {\bibinfo {title} {Ergodicity breaking
  in a model showing many-body localization},\ }\href
  {https://doi.org/10.1209/0295-5075/101/37003} {\bibfield  {journal} {\bibinfo
   {journal} {Europhys. Lett.}\ }\textbf {\bibinfo {volume} {101}},\ \bibinfo
  {pages} {37003} (\bibinfo {year} {2013})}\BibitemShut {NoStop}%
\bibitem [{\citenamefont {Mac\'e}\ \emph {et~al.}(2019)\citenamefont {Mac\'e},
  \citenamefont {Alet},\ and\ \citenamefont
  {Laflorencie}}]{mace2019multifractal}%
  \BibitemOpen
  \bibfield  {author} {\bibinfo {author} {\bibfnamefont {N.}~\bibnamefont
  {Mac\'e}}, \bibinfo {author} {\bibfnamefont {F.}~\bibnamefont {Alet}},\ and\
  \bibinfo {author} {\bibfnamefont {N.}~\bibnamefont {Laflorencie}},\
  }\bibfield  {title} {\bibinfo {title} {Multifractal scalings across the
  many-body localization transition},\ }\href
  {https://doi.org/10.1103/PhysRevLett.123.180601} {\bibfield  {journal}
  {\bibinfo  {journal} {Phys. Rev. Lett.}\ }\textbf {\bibinfo {volume} {123}},\
  \bibinfo {pages} {180601} (\bibinfo {year} {2019})}\BibitemShut {NoStop}%
\bibitem [{\citenamefont {Wei\ss{}e}\ \emph {et~al.}(2006)\citenamefont
  {Wei\ss{}e}, \citenamefont {Wellein}, \citenamefont {Alvermann},\ and\
  \citenamefont {Fehske}}]{weisse2006kernel}%
  \BibitemOpen
  \bibfield  {author} {\bibinfo {author} {\bibfnamefont {A.}~\bibnamefont
  {Wei\ss{}e}}, \bibinfo {author} {\bibfnamefont {G.}~\bibnamefont {Wellein}},
  \bibinfo {author} {\bibfnamefont {A.}~\bibnamefont {Alvermann}},\ and\
  \bibinfo {author} {\bibfnamefont {H.}~\bibnamefont {Fehske}},\ }\bibfield
  {title} {\bibinfo {title} {The kernel polynomial method},\ }\href
  {https://doi.org/10.1103/RevModPhys.78.275} {\bibfield  {journal} {\bibinfo
  {journal} {Rev. Mod. Phys.}\ }\textbf {\bibinfo {volume} {78}},\ \bibinfo
  {pages} {275} (\bibinfo {year} {2006})}\BibitemShut {NoStop}%
\bibitem [{\citenamefont {Mezard}\ \emph {et~al.}(1987)\citenamefont {Mezard},
  \citenamefont {Parisi},\ and\ \citenamefont {Virasoro}}]{mezard1987spin}%
  \BibitemOpen
  \bibfield  {author} {\bibinfo {author} {\bibfnamefont {M.}~\bibnamefont
  {Mezard}}, \bibinfo {author} {\bibfnamefont {G.}~\bibnamefont {Parisi}},\
  and\ \bibinfo {author} {\bibfnamefont {M.}~\bibnamefont {Virasoro}},\ }\href
  {https://books.google.co.uk/books?id=ZIF9QgAACAAJ} {\emph {\bibinfo {title}
  {Spin Glass Theory and Beyond}}},\ Lecture Notes in Physics Series\ (\bibinfo
   {publisher} {World Scientific},\ \bibinfo {year} {1987})\BibitemShut
  {NoStop}%
\bibitem [{\citenamefont {Nahum}\ \emph {et~al.}(2017)\citenamefont {Nahum},
  \citenamefont {Ruhman}, \citenamefont {Vijay},\ and\ \citenamefont
  {Haah}}]{nahum2017quantum}%
  \BibitemOpen
  \bibfield  {author} {\bibinfo {author} {\bibfnamefont {A.}~\bibnamefont
  {Nahum}}, \bibinfo {author} {\bibfnamefont {J.}~\bibnamefont {Ruhman}},
  \bibinfo {author} {\bibfnamefont {S.}~\bibnamefont {Vijay}},\ and\ \bibinfo
  {author} {\bibfnamefont {J.}~\bibnamefont {Haah}},\ }\bibfield  {title}
  {\bibinfo {title} {Quantum entanglement growth under random unitary
  dynamics},\ }\href {https://doi.org/10.1103/PhysRevX.7.031016} {\bibfield
  {journal} {\bibinfo  {journal} {Phys. Rev. X}\ }\textbf {\bibinfo {volume}
  {7}},\ \bibinfo {pages} {031016} (\bibinfo {year} {2017})}\BibitemShut
  {NoStop}%
\bibitem [{\citenamefont {von Keyserlingk}\ \emph {et~al.}(2018)\citenamefont
  {von Keyserlingk}, \citenamefont {Rakovszky}, \citenamefont {Pollmann},\ and\
  \citenamefont {Sondhi}}]{keyserlingk2018operator}%
  \BibitemOpen
  \bibfield  {author} {\bibinfo {author} {\bibfnamefont {C.~W.}\ \bibnamefont
  {von Keyserlingk}}, \bibinfo {author} {\bibfnamefont {T.}~\bibnamefont
  {Rakovszky}}, \bibinfo {author} {\bibfnamefont {F.}~\bibnamefont
  {Pollmann}},\ and\ \bibinfo {author} {\bibfnamefont {S.~L.}\ \bibnamefont
  {Sondhi}},\ }\bibfield  {title} {\bibinfo {title} {Operator hydrodynamics,
  otocs, and entanglement growth in systems without conservation laws},\ }\href
  {https://doi.org/10.1103/PhysRevX.8.021013} {\bibfield  {journal} {\bibinfo
  {journal} {Phys. Rev. X}\ }\textbf {\bibinfo {volume} {8}},\ \bibinfo {pages}
  {021013} (\bibinfo {year} {2018})}\BibitemShut {NoStop}%
\bibitem [{\citenamefont {Chan}\ \emph
  {et~al.}(2018{\natexlab{a}})\citenamefont {Chan}, \citenamefont {De~Luca},\
  and\ \citenamefont {Chalker}}]{chan2018solution}%
  \BibitemOpen
  \bibfield  {author} {\bibinfo {author} {\bibfnamefont {A.}~\bibnamefont
  {Chan}}, \bibinfo {author} {\bibfnamefont {A.}~\bibnamefont {De~Luca}},\ and\
  \bibinfo {author} {\bibfnamefont {J.~T.}\ \bibnamefont {Chalker}},\
  }\bibfield  {title} {\bibinfo {title} {Solution of a minimal model for
  many-body quantum chaos},\ }\href {https://doi.org/10.1103/PhysRevX.8.041019}
  {\bibfield  {journal} {\bibinfo  {journal} {Phys. Rev. X}\ }\textbf {\bibinfo
  {volume} {8}},\ \bibinfo {pages} {041019} (\bibinfo {year}
  {2018}{\natexlab{a}})}\BibitemShut {NoStop}%
\bibitem [{\citenamefont {Rakovszky}\ \emph {et~al.}(2018)\citenamefont
  {Rakovszky}, \citenamefont {Pollmann},\ and\ \citenamefont {von
  Keyserlingk}}]{rakovszky2018diffusive}%
  \BibitemOpen
  \bibfield  {author} {\bibinfo {author} {\bibfnamefont {T.}~\bibnamefont
  {Rakovszky}}, \bibinfo {author} {\bibfnamefont {F.}~\bibnamefont
  {Pollmann}},\ and\ \bibinfo {author} {\bibfnamefont {C.~W.}\ \bibnamefont
  {von Keyserlingk}},\ }\bibfield  {title} {\bibinfo {title} {Diffusive
  hydrodynamics of out-of-time-ordered correlators with charge conservation},\
  }\href {https://doi.org/10.1103/PhysRevX.8.031058} {\bibfield  {journal}
  {\bibinfo  {journal} {Phys. Rev. X}\ }\textbf {\bibinfo {volume} {8}},\
  \bibinfo {pages} {031058} (\bibinfo {year} {2018})}\BibitemShut {NoStop}%
\bibitem [{\citenamefont {Khemani}\ \emph {et~al.}(2018)\citenamefont
  {Khemani}, \citenamefont {Vishwanath},\ and\ \citenamefont
  {Huse}}]{khemani2018operator}%
  \BibitemOpen
  \bibfield  {author} {\bibinfo {author} {\bibfnamefont {V.}~\bibnamefont
  {Khemani}}, \bibinfo {author} {\bibfnamefont {A.}~\bibnamefont
  {Vishwanath}},\ and\ \bibinfo {author} {\bibfnamefont {D.~A.}\ \bibnamefont
  {Huse}},\ }\bibfield  {title} {\bibinfo {title} {Operator spreading and the
  emergence of dissipative hydrodynamics under unitary evolution with
  conservation laws},\ }\href {https://doi.org/10.1103/PhysRevX.8.031057}
  {\bibfield  {journal} {\bibinfo  {journal} {Phys. Rev. X}\ }\textbf {\bibinfo
  {volume} {8}},\ \bibinfo {pages} {031057} (\bibinfo {year}
  {2018})}\BibitemShut {NoStop}%
\bibitem [{\citenamefont {Friedman}\ \emph {et~al.}(2019)\citenamefont
  {Friedman}, \citenamefont {Chan}, \citenamefont {De~Luca},\ and\
  \citenamefont {Chalker}}]{friedman2019spectral}%
  \BibitemOpen
  \bibfield  {author} {\bibinfo {author} {\bibfnamefont {A.~J.}\ \bibnamefont
  {Friedman}}, \bibinfo {author} {\bibfnamefont {A.}~\bibnamefont {Chan}},
  \bibinfo {author} {\bibfnamefont {A.}~\bibnamefont {De~Luca}},\ and\ \bibinfo
  {author} {\bibfnamefont {J.~T.}\ \bibnamefont {Chalker}},\ }\bibfield
  {title} {\bibinfo {title} {Spectral statistics and many-body quantum chaos
  with conserved charge},\ }\href
  {https://doi.org/10.1103/PhysRevLett.123.210603} {\bibfield  {journal}
  {\bibinfo  {journal} {Phys. Rev. Lett.}\ }\textbf {\bibinfo {volume} {123}},\
  \bibinfo {pages} {210603} (\bibinfo {year} {2019})}\BibitemShut {NoStop}%
\bibitem [{\citenamefont {Chan}\ \emph
  {et~al.}(2018{\natexlab{b}})\citenamefont {Chan}, \citenamefont {De~Luca},\
  and\ \citenamefont {Chalker}}]{chan2018spectral}%
  \BibitemOpen
  \bibfield  {author} {\bibinfo {author} {\bibfnamefont {A.}~\bibnamefont
  {Chan}}, \bibinfo {author} {\bibfnamefont {A.}~\bibnamefont {De~Luca}},\ and\
  \bibinfo {author} {\bibfnamefont {J.~T.}\ \bibnamefont {Chalker}},\
  }\bibfield  {title} {\bibinfo {title} {Spectral statistics in spatially
  extended chaotic quantum many-body systems},\ }\href
  {https://doi.org/10.1103/PhysRevLett.121.060601} {\bibfield  {journal}
  {\bibinfo  {journal} {Phys. Rev. Lett.}\ }\textbf {\bibinfo {volume} {121}},\
  \bibinfo {pages} {060601} (\bibinfo {year} {2018}{\natexlab{b}})}\BibitemShut
  {NoStop}%
\bibitem [{\citenamefont {Lazarides}\ \emph
  {et~al.}(2014{\natexlab{a}})\citenamefont {Lazarides}, \citenamefont {Das},\
  and\ \citenamefont {Moessner}}]{lazarides2014equilibrium}%
  \BibitemOpen
  \bibfield  {author} {\bibinfo {author} {\bibfnamefont {A.}~\bibnamefont
  {Lazarides}}, \bibinfo {author} {\bibfnamefont {A.}~\bibnamefont {Das}},\
  and\ \bibinfo {author} {\bibfnamefont {R.}~\bibnamefont {Moessner}},\
  }\bibfield  {title} {\bibinfo {title} {Equilibrium states of generic quantum
  systems subject to periodic driving},\ }\href
  {https://doi.org/10.1103/PhysRevE.90.012110} {\bibfield  {journal} {\bibinfo
  {journal} {Phys. Rev. E}\ }\textbf {\bibinfo {volume} {90}},\ \bibinfo
  {pages} {012110} (\bibinfo {year} {2014}{\natexlab{a}})}\BibitemShut
  {NoStop}%
\bibitem [{\citenamefont {D'Alessio}\ and\ \citenamefont
  {Rigol}(2014)}]{dalessio2014long}%
  \BibitemOpen
  \bibfield  {author} {\bibinfo {author} {\bibfnamefont {L.}~\bibnamefont
  {D'Alessio}}\ and\ \bibinfo {author} {\bibfnamefont {M.}~\bibnamefont
  {Rigol}},\ }\bibfield  {title} {\bibinfo {title} {Long-time behavior of
  isolated periodically driven interacting lattice systems},\ }\href
  {https://doi.org/10.1103/PhysRevX.4.041048} {\bibfield  {journal} {\bibinfo
  {journal} {Phys. Rev. X}\ }\textbf {\bibinfo {volume} {4}},\ \bibinfo {pages}
  {041048} (\bibinfo {year} {2014})}\BibitemShut {NoStop}%
\bibitem [{\citenamefont {Ponte}\ \emph
  {et~al.}(2015{\natexlab{a}})\citenamefont {Ponte}, \citenamefont {Chandran},
  \citenamefont {Papi{\'c}},\ and\ \citenamefont
  {Abanin}}]{ponte2015periodically}%
  \BibitemOpen
  \bibfield  {author} {\bibinfo {author} {\bibfnamefont {P.}~\bibnamefont
  {Ponte}}, \bibinfo {author} {\bibfnamefont {A.}~\bibnamefont {Chandran}},
  \bibinfo {author} {\bibfnamefont {Z.}~\bibnamefont {Papi{\'c}}},\ and\
  \bibinfo {author} {\bibfnamefont {D.~A.}\ \bibnamefont {Abanin}},\ }\bibfield
   {title} {\bibinfo {title} {Periodically driven ergodic and many-body
  localized quantum systems},\ }\href
  {https://www.sciencedirect.com/science/article/pii/S0003491614003212}
  {\bibfield  {journal} {\bibinfo  {journal} {Annals of Physics}\ }\textbf
  {\bibinfo {volume} {353}},\ \bibinfo {pages} {196} (\bibinfo {year}
  {2015}{\natexlab{a}})}\BibitemShut {NoStop}%
\bibitem [{\citenamefont {Lazarides}\ \emph
  {et~al.}(2014{\natexlab{b}})\citenamefont {Lazarides}, \citenamefont {Das},\
  and\ \citenamefont {Moessner}}]{lazarides2014periodic}%
  \BibitemOpen
  \bibfield  {author} {\bibinfo {author} {\bibfnamefont {A.}~\bibnamefont
  {Lazarides}}, \bibinfo {author} {\bibfnamefont {A.}~\bibnamefont {Das}},\
  and\ \bibinfo {author} {\bibfnamefont {R.}~\bibnamefont {Moessner}},\
  }\bibfield  {title} {\bibinfo {title} {Periodic thermodynamics of isolated
  quantum systems},\ }\href {https://doi.org/10.1103/PhysRevLett.112.150401}
  {\bibfield  {journal} {\bibinfo  {journal} {Phys. Rev. Lett.}\ }\textbf
  {\bibinfo {volume} {112}},\ \bibinfo {pages} {150401} (\bibinfo {year}
  {2014}{\natexlab{b}})}\BibitemShut {NoStop}%
\bibitem [{\citenamefont {Lazarides}\ \emph {et~al.}(2015)\citenamefont
  {Lazarides}, \citenamefont {Das},\ and\ \citenamefont
  {Moessner}}]{lazarides2015fate}%
  \BibitemOpen
  \bibfield  {author} {\bibinfo {author} {\bibfnamefont {A.}~\bibnamefont
  {Lazarides}}, \bibinfo {author} {\bibfnamefont {A.}~\bibnamefont {Das}},\
  and\ \bibinfo {author} {\bibfnamefont {R.}~\bibnamefont {Moessner}},\
  }\bibfield  {title} {\bibinfo {title} {Fate of many-body localization under
  periodic driving},\ }\href {https://doi.org/10.1103/PhysRevLett.115.030402}
  {\bibfield  {journal} {\bibinfo  {journal} {Phys. Rev. Lett.}\ }\textbf
  {\bibinfo {volume} {115}},\ \bibinfo {pages} {030402} (\bibinfo {year}
  {2015})}\BibitemShut {NoStop}%
\bibitem [{\citenamefont {Ponte}\ \emph
  {et~al.}(2015{\natexlab{b}})\citenamefont {Ponte}, \citenamefont
  {Papi\ifmmode~\acute{c}\else \'{c}\fi{}}, \citenamefont {Huveneers},\ and\
  \citenamefont {Abanin}}]{ponte2015many}%
  \BibitemOpen
  \bibfield  {author} {\bibinfo {author} {\bibfnamefont {P.}~\bibnamefont
  {Ponte}}, \bibinfo {author} {\bibfnamefont {Z.}~\bibnamefont
  {Papi\ifmmode~\acute{c}\else \'{c}\fi{}}}, \bibinfo {author} {\bibfnamefont
  {F.}~\bibnamefont {Huveneers}},\ and\ \bibinfo {author} {\bibfnamefont
  {D.~A.}\ \bibnamefont {Abanin}},\ }\bibfield  {title} {\bibinfo {title}
  {Many-body localization in periodically driven systems},\ }\href
  {https://doi.org/10.1103/PhysRevLett.114.140401} {\bibfield  {journal}
  {\bibinfo  {journal} {Phys. Rev. Lett.}\ }\textbf {\bibinfo {volume} {114}},\
  \bibinfo {pages} {140401} (\bibinfo {year} {2015}{\natexlab{b}})}\BibitemShut
  {NoStop}%
\bibitem [{\citenamefont {Mukherjee}\ \emph {et~al.}(2019)\citenamefont
  {Mukherjee}, \citenamefont {Nandy}, \citenamefont {Sen},\ and\ \citenamefont
  {Sengupta}}]{mukherjee2019collapse}%
  \BibitemOpen
  \bibfield  {author} {\bibinfo {author} {\bibfnamefont {B.}~\bibnamefont
  {Mukherjee}}, \bibinfo {author} {\bibfnamefont {S.}~\bibnamefont {Nandy}},
  \bibinfo {author} {\bibfnamefont {A.}~\bibnamefont {Sen}},\ and\ \bibinfo
  {author} {\bibfnamefont {K.}~\bibnamefont {Sengupta}},\ }\bibfield  {title}
  {\bibinfo {title} {Collapse and revival of quantum many-body scars via
  floquet engineering},\ }\href {https://arxiv.org/abs/1907.08212} {\bibfield
  {journal} {\bibinfo  {journal} {arXiv preprint arXiv:1907.08212}\ } (\bibinfo
  {year} {2019})}\BibitemShut {NoStop}%
\bibitem [{\citenamefont {Haldar}\ \emph {et~al.}(2019)\citenamefont {Haldar},
  \citenamefont {Sen}, \citenamefont {Moessner},\ and\ \citenamefont
  {Das}}]{haldar2019scars}%
  \BibitemOpen
  \bibfield  {author} {\bibinfo {author} {\bibfnamefont {A.}~\bibnamefont
  {Haldar}}, \bibinfo {author} {\bibfnamefont {D.}~\bibnamefont {Sen}},
  \bibinfo {author} {\bibfnamefont {R.}~\bibnamefont {Moessner}},\ and\
  \bibinfo {author} {\bibfnamefont {A.}~\bibnamefont {Das}},\ }\bibfield
  {title} {\bibinfo {title} {Scars in strongly driven floquet matter: resonance
  vs emergent conservation laws},\ }\href {https://arxiv.org/abs/1909.04064}
  {\bibfield  {journal} {\bibinfo  {journal} {arXiv preprint arXiv:1909.04064}\
  } (\bibinfo {year} {2019})}\BibitemShut {NoStop}%
\end{thebibliography}%

\end{document}